\documentclass[conference,onecolumn,12pt]{IEEEtran}
\usepackage{verbatim,epsfig,times,subfigure}
\usepackage{amsmath,amsfonts,amssymb,mathrsfs}
\pdfoutput=1
\usepackage[pdftex]{color}
\usepackage{url}
\usepackage{hyperref}
\definecolor{brown}{cmyk}{0,0.81,1,0.60}
\definecolor{magenta}{rgb}{0.4,0.7,0}
\definecolor{gray}{rgb}{0.5,0.5,0.5}
\definecolor{red}{rgb}{1,0,0}
\definecolor{green}{rgb}{0.5,0,0.5}
\definecolor{blue}{rgb}{0,0,1}

\newtheorem{definition}{{\bf \hspace{-0.18in} Definition}}
\newtheorem{assumption}{{\bf \hspace{-0.18in} Assumption}}
\newtheorem{theorem}{{\bf \hspace{-0.18in} Theorem}}
\newtheorem{lemma}{{\bf \hspace{-0.18in} Lemma}}
\newtheorem{axiom}{{\bf \hspace{-0.18in} Axiom}}

\newtheorem{corollary}{{\bf \hspace{-0.18in} Corollary}}
\def\done{\hspace*{\fill} \rule{1.8mm}{2.5mm}}

\begin{document}

\title{The Public Option: a Non-regulatory Alternative to Network
  Neutrality  }

\author{\IEEEauthorblockN{Richard T. B. Ma}
\IEEEauthorblockA{Department of Computer Science\\
National University of Singapore\\
Email: tbma@comp.nus.edu.sg}
\and
\IEEEauthorblockN{Vishal Misra}
\IEEEauthorblockA{Department of Computer Science\\
Columbia University,
New York, USA\\
Email: misra@cs.columbia.edu}
}

\maketitle


\begin{abstract}
Network neutrality and the role of regulation on the Internet have been heavily debated in recent times. Amongst the various definitions
of network neutrality, we focus on the one which \emph{prohibits paid prioritization of content} and we present an analytical treatment of the topic. We develop a model of the Internet ecosystem in terms of three primary players: consumers, ISPs and content providers.
Our analysis looks at this issue from the point of view of the consumer, and we describe the desired state of the system as one which maximizes  {\em consumer surplus}. By analyzing different scenarios of monopoly and competition, we obtain different conclusions on the desirability of regulation. We also introduce the notion of a \emph{Public Option ISP}, an ISP that carries traffic in a network neutral manner. Our major findings are (i) in a monopolistic scenario, network neutral regulations benefit consumers; however, the introduction of a Public Option ISP is even better for consumers, as it aligns the interests of the monopolistic ISP with the consumer surplus and (ii) in an oligopolistic situation, the presence of a Public Option ISP is again preferable to network neutral regulations, although the presence of competing price-discriminating ISPs provides the most desirable situation for the consumers.
\end{abstract}

\section{Introduction}
Since around 2005, network neutrality has been a hotly debated topic amongst law and policy makers. The core debate has centered around the argument whether ISPs should be allowed to provide service differentiation and/or user discrimination, with the notion of ``user'' being either content providers (CPs) or consumers. Proponents of network neutrality, mostly the CPs, have argued that the Internet has been ``neutral'' since its inception and that has been a critical factor in the innovation and rapid growth that has happened on it. Opponents of network neutrality, mostly the ISPs, claim that without some sort of service differentiation, ISPs will lose the incentive to invest in the
networks and  the end user experience will suffer. Both camps implicitly or explicitly claim that their approach is beneficial for consumers. A recent Federal Communications Commission (FCC) vote \cite{fcc-vote} in the US has sided with the proponents, although the ruling leaves some room for service differentiation in wireless networks. The controvesy rages on though with corporations like Verizon filing lawsuits challenging the ruling and a ``toll-tax'' dispute between Level3/Netflix and Comcast being cast as a network neutrality issue.

We study the issue explicitly from the point of view of the consumer under both  monopolistic and  oligopolistic scenarios.
A lot of arguments against network neutrality live in an idealized world where economies of scale do not exist and monopolies cannot emerge, and therefore perfect competition solves all problems. We believe reality is more nuanced and hence we examine monopolistic scenarios as well.

Our approach to the analysis is a game theoretic one, and we focus on the consumer surplus. We model the rate allocation mechanism of the system and the user demand for different CPs. The interplay between the two determines the rate equilibrium for all traffic flows.
Our model of price discrimination is for the ISPs to offer two classes of service to CPs. The ISP divides its capacity into a premium and ordinary class, and CPs get charged for carrying traffic in the premium class, and more details are presented in Section~\ref{sec:differentiation}. We then identify and analyze the strategic games played between ISPs, CPs and consumers in Section~\ref{sec:monopoly} for a monopolistic scenario and in Section~\ref{sec:oligopoly} for oligopolistic scenarios. 
In Section~\ref{sec:public-option}, we introduce the notion of a \emph{Public Option} ISP 
 which is neutral to all CPs. The Public Option ISP can be implemented by processes like local loop unbundling~\cite{unbundling-wikipedia} in a monopolistic market and either government or a private organization can run the ISP and still be profitable \cite{dovrolis08can}.
  Given the framework,  our major findings are: 

\begin{itemize}
\item The impact of network neutrality on consumer surplus depends on the nature of competition at the ISP level. Concretely, a neutral
  network is beneficial for consumer surplus under a monopolistic regime (Section \ref{sec:monopoly}), whereas a non-neutral network is advantageous for consumers under an oligopolistic scenario (Section \ref{sec:oligopoly}).
\item The capacity of the network plays a big role on the regulations for consumer surplus. Price discrimination in a monopolistic network where capacity is plentiful can have a damaging effect on consumer surplus, whereas it might gain higher utility for the consumer in a network where capacity is extremely scarce (Section \ref{sec:monopoly}).
\item Introducing a Public Option ISP is advantageous for consumers. 
  In a monopolistic situation, the Public Option ISP offers the best scenario for consumers (Theorem \ref{theorem:optimality_two_ISPs}), followed by network neutral regulations, and an unregulated market being the worst.
\item In an oligopolistic situation, the Public Option ISP is still preferable to network neutral regulations; however, since the incentive for an ISP to gain market share is aligned with the maximization of consumer surplus (Theorem \ref{theorem:multiISP_best_response}), no regulation is needed to protect the consumers.
\item Under an oligopolistic competition, any ISP's optimal pricing
  and service differentiation strategy, whether network neutral or not, will be close the one that maximizes consumer surplus (Theorem \ref{theorem:multiISP_best_response} and Corollary \ref{corollary:epsilon_Nash}). Moreover, under a probable equilibrium where ISPs use homogenous strategies, their market shares will be proportional to their capacities (Lemma \ref{lemma:market_share}), which implies that ISPs do have incentives to invest and expand capacity so as to their increase market shares.
\end{itemize}

Our paper sheds new light on the network neutrality debate and concretely identifies where and how regulation can help. Additionally, our identification of the Public Option ISP is especially important as it provides a solution that combines the best of both worlds, protecting consumer interests without enforcing strict regulations on all ISPs. We start with describing our model in the next section.

\section{Three-party Ecosystem Model}\label{sec:model}
We consider a model of the Internet with three parties: 1) CPs, 2) ISPs and 3) consumers.
We focus on a fixed consumer group in a targeted geographic region. We
denote $M$ as the number of consumers in the region\footnote{Note that
  $M$ can also be interpreted as the \emph{average} or \emph{peak}
  number of consumers accessing the Internet simultaneously in the
  region, which will scale with the total number of actual consumers
  in a region. This does not change the nature of any of the results
  we describe subsequently, but gives a more realistic interpretation
  of the rate equilibrium.}. Each consumer subscribes to an Internet access service via an ISP. We consider the scenarios where one monopolistic ISP $\mathit I$ or a set $\cal I$ of competing oligopolistic ISPs  provide the Internet access for the consumers. We denote $\cal N$ as the set of CPs from which the consumers request content. We define $N=|\cal N|$ as the number of CPs. Our model does not include the backbone ISPs for two reasons. First, the bottleneck of the Internet is often at the last-mile connection towards the consumers \cite{weber-pricing}, both wired and wireless.
We focus on the regional or so-called {\em eyeball ISPs} \cite{clark07} that provide the bottleneck last-mile towards the consumers. Second, the recent concern on network neutrality manifests itself in the cases where the last-mile ISPs, e.g. France Telecom, Telecom Italia and Vodafone, intended to differentiate services and charge CPs, e.g. Apple and Google, for service fees \cite{apple10Bloomberg}.

We denote $\mu$ as the last-mile bottleneck capacity towards the consumers in the region.
\begin{figure}[ht]
\centering
\includegraphics[width=3.8in, angle=0]{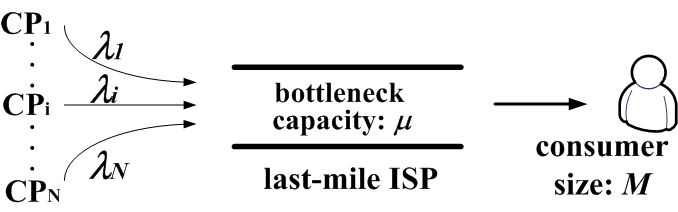}
\caption{Contention at the last-mile bottleneck link.}
\label{figure:equ_rate2}
\end{figure}
Figure \ref{figure:equ_rate2} depicts the contention at the bottleneck among different flows from the CPs.
We denote $\lambda_i$ as the aggregate throughput rate from CP $i$ to the consumers. Because consumers initiate downloads and retrieve content from the CPs, we first model the consumer demands so as to characterize the CPs' throughput rates $\lambda_i$s. 
Given a set $\cal N$ of CPs, a group of $M$ consumers and a link with capacity $\mu$, we denote the system as a triple $(M,\mu,{\cal N})$.

\subsection{Throughput Demand}
We denote $\hat{\theta}_i$ as the {\em unconstrained throughput} for a typical user of CP $i$. For instance, the unconstrained throughput for the highest quality Netflix streaming movie is about 5 Mbps~\cite{netflix-techblog}, and given an average query page of 20 KB and an average query response time of .25 seconds~\cite{google-corporate}, the unconstrained throughput for a Google search is about 600 Kbps, or just over $1/10^{th}$ of Netflix.
We denote $\alpha_i \in (0,1]$ as the percentage of consumers that ever access CP $i$'s content, which  models the popularity of the content of CP $i$. We define $\hat\lambda_i = \alpha_i M\hat\theta_i$ as the unconstrained throughput of CP $i$. Without contention, CP $i$'s throughput $\lambda_i$ equals $\hat\lambda_i$. However, when the capacity $\mu$ cannot support the unconstrained throughput from all CPs, i.e. $\mu < \sum_{i\in \cal N} \hat\lambda_i$, two things will happen: 1) a typical user of CP $i$ obtains throughput $\theta_i < \hat\theta_i$ from CP $i$, and 2) some active users might stop downloading content from CP $i$ when $\theta_i$ goes below certain threshold, e.g. users of real-time content like Netflix. We denote $\theta_i$ as the {\em achievable throughput} for the consumers downloading content from CP $i$.
We define a demand function $d_i(\theta_i)$ which represents the percentage of consumers that still demand content from CP $i$ under the achievable throughput $\theta_i$.

\vspace{-0.02in}
{\assumption For any CP $i$, the demand $d_i(\cdot)$ is a non-negative, continuous and non-decreasing function defined on the domain of  $[0,\hat\theta_i]$, and satisfies $d_i(\hat\theta_i)=1$.
\label{assumption:demand}}

We define the aggregate throughput as $\lambda_{\cal N} = \sum_{i\in\cal N} \lambda_i$, where each CP $i$'s throughput $\lambda_i$ is defined as follows:
\begin{equation}
\lambda_i(\theta_i)  = \alpha_i M d_i(\theta_i)\theta_i.
\end{equation}

\subsection{Rate Allocation Mechanism}
When multiple flows share the same bottleneck link, they compete for capacity.
The rates allocated to the flows depend on the rate allocation mechanism being use in the system.
{\definition  A {\em rate allocation mechanism} is a function that maps any fixed demand profile $\{d_i:i\in \cal N\}$\footnote{Without a bracket, we use $d_i$ as a fixed demand.} to
an achievable throughput profile $\{\theta_i: i\in \cal N\}$. \label{definition:rate_allocation}}

\vspace{0.05in}
A rate allocation mechanism can be a flow control mechanism, e.g. CBR and VBR mechanisms, under which the bottleneck link decides the rates for each flow in a centralized manner, or a window-based end-to-end congestion control mechanism, e.g. TCP, under which each flow maintains a sliding window and adapts the its size based on implicit feedback from the network, e.g. the round-trip delay. We consider generic rate allocation mechanisms and assume that the resulting rate allocation obeys the physical constraints of the system and satisfies some desirable properties.
{\axiom $\theta_i \leq \hat \theta_i \ $ for all $i\in \cal N$.
\label{axiom:bound1}}
{\axiom $\lambda_{\cal N} = \min \big\{ \mu, \ \sum_{i\in \cal N}  \hat\lambda_i  \big\}$.\label{axiom:bound2}}

\vspace{0.05in}
\noindent The above axioms characterize the feasibility of an allocation: the aggregate rate cannot exceed the capacity and the individual rate would not exceed its unconstrained throughput. It also characterizes a {\em work-conserving} property: if congestion can be alleviated without increasing capacity $\mu$, the allocation would do so by fully utilizing the capacity.


{\axiom[Monotonicity] A rate allocation is {\em monotonic} if
for any $M>0$ and capacity $\mu_1<\mu_2$, the achievable throughput $\theta_i$ for any $i\in \cal N$ satisfies
\[ \theta_i(M,\mu_2,{\cal N}) \geq  \theta_i(M,\mu_1,{\cal N}). \]
\label{axiom:monotonicity}}
The Monotonicity axiom implies that if a flow gets an achievable throughput in one system, it will be allocated at least that amount of throughput under a less congested system.

{\axiom[Independence of Scale] A rate allocation is {\em independent of scale} if
for any $\xi>0$, the achievable throughput $\theta_i$ for any $i\in \cal N$ satisfies
\[\theta_i(M,\mu,{\cal N}) = \theta_i(\xi M,\xi \mu,{\cal N}).\]
\label{axiom:linear_scale}}
The Independence of Scale axiom states that if the capacity scales at the same rate as the consumer size, each flow's achievable throughput $\theta_i$ remains the same.

{\assumption The network system implements a rate allocation mechanism that satisfies Axiom \ref{axiom:bound1} to \ref{axiom:linear_scale}. \label{assumption:regular_mechanism}}

\subsection{Rate Equilibrium}
The demand functions map the achievable throughput to a level of demand; the rate allocation mechanisms map fixed demands to achievable throughput.
The interplay between a rate allocation mechanism and the demand functions determines the system rate equilibrium
as the following theorem.

{\theorem A system $(M,\mu,\cal N)$ has a unique rate equilibrium $\{\theta_i:i\in\cal N\}$ under Assumption \ref{assumption:demand} and Axiom \ref{axiom:bound1} to \ref{axiom:monotonicity}.
\label{theorem:unique_rate_equilibrium}}

\noindent {\bf Proof of Theorem \ref{theorem:unique_rate_equilibrium}:}
Based on Assumption \ref{assumption:demand}, we know that for any $i \in \cal N$, the throughput $\lambda_i(\theta_i)$ is a non-decreasing and continuous function. By Axiom \ref{axiom:bound1}, $\lambda_i(\theta_i)$ has a range of $[0,\hat\lambda_i]$.
We show the uniqueness of the rate equilibrium by the following two cases.

We first consider the case where $\mu\geq \sum_{i\in\cal N}\hat\lambda_i$. By Axiom \ref{axiom:bound2}, $\lambda_{\cal N} = \sum_{i\in\cal N}\lambda_i(\theta_i) = \sum_{i\in\cal N}\hat\lambda_i$. Because $\lambda_i \leq \hat \lambda_i$ for all $i\in\cal N$, we must have $\lambda_i = \hat \lambda_i$. Therefore, the unique rate equilibrium must be $\{\theta_i = \hat \theta_i:i\in\cal N\}$.

We then consider the case where $\mu< \sum_{i\in\cal N}\hat\lambda_i$. By Axiom \ref{axiom:bound2}, $\lambda_{\cal N} = \sum_{i\in\cal N}\lambda_i(\theta_i) = \mu$. Because each $\lambda_i(\theta_i)$ is non-decreasing in $\theta_i$ and continuous in the range of $[0,\hat\lambda_i]$, we can always find a solution $\{\theta_i:i\in\cal N\}$ that satisfies $\sum_{i\in\cal N}\lambda_i(\theta_i) = \mu$. Suppose there exists two equilibrium solutions $\{\theta_i^1:i\in\cal N\}$ and $\{\theta_i^2:i\in\cal N\}$. For any $\varepsilon >0$, we denote $\{\theta_i^3:i\in\cal N\}$ as a rate equilibrium of the system $(M,\mu-\varepsilon,\cal N)$.  By Axiom \ref{axiom:bound2}, $\sum_{i\in\cal N}\lambda_i(\theta_i^3) = \mu-\varepsilon$. By Axiom \ref{axiom:monotonicity}, $\theta_i^3 \leq \min \{\theta_i^1,\theta_i^2\}$ for all $i\in\cal N$. Thus, we have
\begin{IEEEeqnarray*}{c}
\mu-\varepsilon = \sum_{i\in\cal N}\lambda_i(\theta_i^3) \leq  \sum_{i\in\cal N}\lambda_i(\min \{\theta_i^1,\theta_i^2\}) \\
= \sum_{i\in\cal N} \min \{ \lambda_i(\theta_i^1),\lambda_i(\theta_i^2)\} <\mu.
\end{IEEEeqnarray*}
However, the above inequality would be violated when $\varepsilon$ approaches zero. By showing this contradiction, we prove the uniqueness of the rate equilibrium under $\mu< \sum_{i\in\cal N}\hat\lambda_i$.
\done

\bigskip

We define $\nu = \mu / M$ as the per capita capacity of the system. By Axiom \ref{axiom:linear_scale},
we further characterize the rate equilibrium  $\{\theta_i:i\in\cal N\}$ as follows.

{\lemma Under Assumption \ref{assumption:demand} and \ref{assumption:regular_mechanism},  for all $i \in \cal N$,
$\theta_i$ in equilibrium can be expressed as $\theta_i(M,\mu,{\cal N})=\theta_i(\nu,{\cal N})$, which is a non-decreasing and continuous function in $\nu$.
\label{lemma:theta}}

\noindent {\bf Proof of Lemma \ref{lemma:theta}:}
By Axiom \ref{axiom:linear_scale}, $\theta_i(M_1,\mu_1,\cal N)$ equals $\theta_i(M_2,\mu_2,\cal N)$ if $M_1/\mu_1 = M_2/\mu_2$. Therefore, we can denote $\theta_i(\nu,{\cal N}) = \theta_i(M,\mu,\cal N)$ for all systems $(M,\mu,\cal N)$ with $\nu = M/\mu$. Given any two systems $(M_1,\mu_1,\cal N)$ and $(M_2,\mu_2,\cal N)$ with $\nu_1 \leq \nu_2$, we first  show that $\theta_i(\nu_1,{\cal N}) \leq \theta_i(\nu_2,{\cal N})$ for all $i\in \cal N$. Suppose there exists $i \in \cal N$ such that $\theta_i(\nu_1,{\cal N}) > \theta_i(\nu_2,{\cal N})$. By Axiom \ref{axiom:linear_scale}, we have
\begin{IEEEeqnarray*}{c}
\theta_i(M_1,\mu_1,{\cal N}) = \theta_i(\nu_1,{\cal N}) \\ > \theta_i(\nu_2,{\cal N}) = \theta_i(M_2,\mu_2,{\cal N}) = \theta_i(M_1,\frac{M_1}{M_2}\mu_2,{\cal N}).
\end{IEEEeqnarray*}
Because $\frac{M_1}{M_2}\mu_2 > \mu_1$, the above inequality violates Axiom \ref{axiom:monotonicity}. By showing this contradiction, we show that $\theta_i(\nu,\cal N)$ is a non-decreasing function of $\nu$.
As a result, if $\theta_i(\nu,\cal N)$ is not continuous, then it can only have upward jumps. Therefore, $\lambda_{\cal N} = \sum_{i\in\cal N} \lambda_i(\theta_i)$ will only have upward jumps as well. However, by Axiom \ref{axiom:bound2}, $\lambda_{\cal N}$ has to meet the equality constraint $\lambda_{\cal N} = \min \big\{ \mu, \ \sum_{i\in \cal N}  \hat\lambda_i  \big\}$, which implies that when $\nu$ increases from zero, $\lambda_{\cal N}$ cannot have jumps. Therefore, each $\theta_i$ has to be continuous in $\nu$.
\done

\bigskip

Lemma \ref{lemma:theta} states that when $\nu$ increases, users'
achievable throughput $\theta_i$ would not worse off for any CP $i$.

We denote $\phi_i \geq 0$ as the per unit traffic utility that the consumers obtain by receiving content from CP $i$. This utility can be derived from communicating with friends, e.g. Skype, watching movies, e.g. Netflix, obtaining information, e.g. Google, RapidShare, and etc. Our model does not assume any form of this utility. We denote $CS$ as the consumer surplus defined as  $CS = \sum_{i\in \cal N} \phi_i \lambda_i $ and denote $\Phi$ as the per capita consumer surplus defined as
\begin{IEEEeqnarray}{c}
\Phi = \frac{CS}{M} = \frac{1}{M}\sum_{i\in \cal N} \phi_i \lambda_i(\theta_i) = \sum_{i\in \cal N} \phi_i \alpha_i d_i(\theta_i)\theta_i. 
\end{IEEEeqnarray}

{\theorem Under Assumption \ref{assumption:demand} and \ref{assumption:regular_mechanism},
the per capita consumer surplus $\Phi$ can be expressed as $\Phi(M,\mu,{\cal N}) = \Phi(\nu,\cal N)$, which is non-decreasing function in $\nu$. In particular, it strictly increases in $\nu\in[0, \sum_{i\in \cal N}\alpha_i\hat\theta_i]$.
\label{theorem:Phi_monotonicity}}

Theorem \ref{theorem:Phi_monotonicity} states that the per capita consumer surplus will strictly increase with the system per capita capacity $\nu$, unless it is already maximized when unconstrained throughput is obtained. This result does not depend on the values of $\phi_i$s.

\noindent {\bf Proof of Theorem \ref{theorem:Phi_monotonicity}:} If we define $\Phi_i = \phi_i \alpha_i d_i(\theta_i)\theta_i$ for all $i\in\cal N$, by definition $\Phi = \sum_{i\in \cal N} \Phi_i$. Since $\theta_i$ is a function of $\nu$ by Lemma \ref{lemma:theta}, $\Phi_i$ and $\Phi$ are functions of $\nu$ as well.
By Assumption \ref{assumption:demand}, $\Phi_i$ is a non-decreasing function of $\theta_i$. By Lemma \ref{lemma:theta}, $\theta_i$ is non-decreasing in $\nu$; therefore, $\Phi_i$ and then $\Phi$ are non-decreasing in $\nu$.
By Axiom \ref{axiom:bound2}, $\lambda_{\cal N} = \mu$ when $\nu \in [0,\sum_{i\in\cal N}\alpha_i\hat\theta_i]$, which implies that when $\nu$ increases, there must exist some $i\in\cal N$ with $\theta_i$ strictly increasing. As a result, $\Phi_i$ and $\Phi$ have to be strictly increasing as well.
\done

\bigskip

\subsection{Examples and Illustrations}
In this subsection, we illustrate some examples of demand functions and rate allocation mechanisms. 
\subsubsection{Demand as a function of throughput sensitivity}
Distinct CPs often have different demand patterns. For example, the demand for real-time applications decreases dramatically when its throughput drops below certain threshold where performance cannot be tolerated by most of users. We can characterize this throughput sensitivity of the CPs by a positive parameter $\beta_i$ and consider the demand function
\begin{equation}
d_i (\theta_i) = \displaystyle{ e^{-\beta_i\big(\frac{\hat \theta_i}{\theta_i} - 1\big)} =  e^{-\beta_i\big(\frac{1}{\omega_i} - 1\big)} },
\label{equation:demand_function}
\end{equation}
where we define $\omega_i = \theta_i / \hat \theta_i$ as the percentage of unconstrained throughput achieved for CP $i$. The user demand decays exponentially with the level of congestion (measured by $\frac{\hat\theta_i - \theta_i}{\theta_i}$, the ratio of unsatisfied and achieved throughput) scaled by $\beta_i$.
This demand function 
distinguishes the CPs via their throughput sensitivity $\beta_i$: larger $\beta_i$ indicates higher sensitivity to throughput for CP $i$.
\begin{figure}[h]
\centering
\includegraphics[width=4in, angle=0]{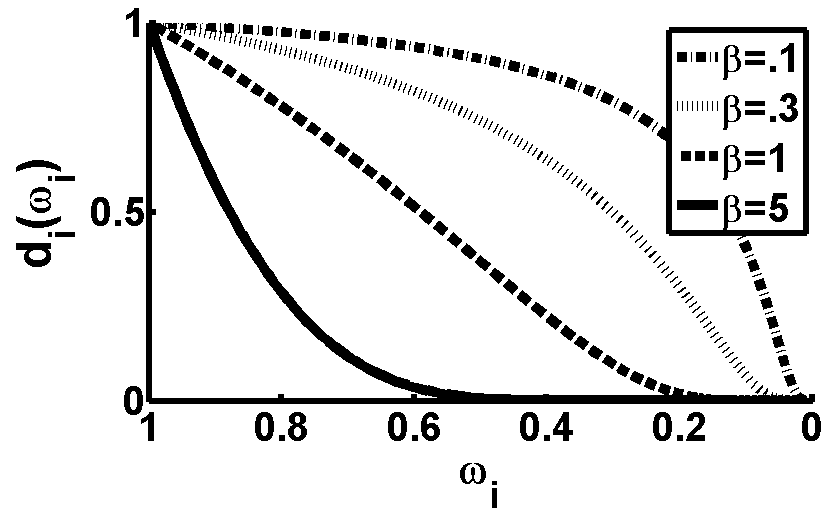}
\caption{Demand function $d_i(\omega_i)$.}
\label{figure:plot_di}
\end{figure}
Figure \ref{figure:plot_di} illustrates the demand functions with various values of $\beta_i$. To normalize $\hat\theta_i$, we plot $d_i$ against $\omega_i$ instead of $\theta_i$.
We observe that when throughput drops linearly, the demand drops sharply for large $\beta_i$, e.g. when $\beta_i=5$, the demand is halved with a $10\%$ drop in throughput from $\hat\theta_i$. Large $\beta_i$s can be used to model CPs that have stringent throughput requirements, e.g. Netflix; while, small $\beta_i$s can be used to model CPs that are less sensitive to throughput, e.g. a Google search query.

\subsubsection{End-to-end congestion control mechanisms}

Due to the end-to-end design principle of the Internet \cite{clark88design}, congestion control has been implemented by window-based protocols, i.e. TCP and its variations. Mo and Walrand \cite{mo00fair} showed that a class of $\alpha$-proportional fair solutions\footnote{Any $\alpha$-proportional fair solution also satisfies Assumption \ref{assumption:regular_mechanism}.} can be implemented by window-based end-to-end protocols. Among the class of $\alpha$-proportional fair solutions, the max-min fair allocation, a special case with $\alpha = \infty$, is the result of the AIMD mechanism of TCP \cite{dmchiu}. 
Differing round trip times, receiver window sizes and loss rates can result in different bandwidths, but to a first approximation, TCP provides a max-min fair allocation of available bandwidth amongst flows. Although other protocols, e.g. UDP, coexist in the Internet, recent research  \cite{craig10internet} sees a growing concentration of application traffic, especially video, over TCP.

We illustrate the rate allocation under the max-min fair mechanism using an example of three CPs with demand functions of Equation (\ref{equation:demand_function}) and parameters $(\alpha_1,\hat\theta_1,\beta_1)=(1,1,0.1)$, $(\alpha_2,\hat\theta_2,\beta_2)=(0.3,10,3)$ and $(\alpha_3,\hat\theta_3,\beta_3)=(0.5,3,5)$. CP $1$ represents Google-type of CPs that are extensively accessed and less sensitive to throughput. CP $2$ represents Netflix-type of CPs that are more throughput-sensitive and have high unconstrained throughput. CP $3$ represents Skype-type of CPs that are extremely sensitive to throughput and have medium unconstrained throughput.

\begin{figure}[ht]
\centering
\includegraphics[width=3.2in, angle=0]{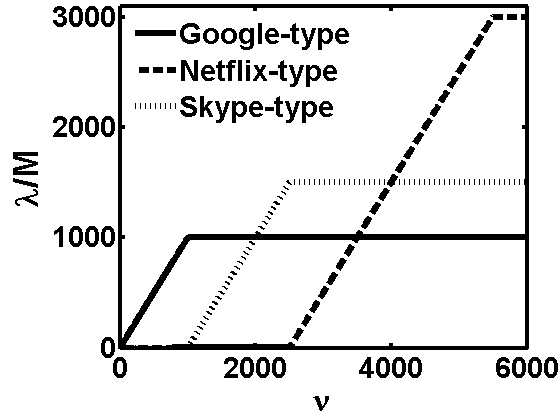}
\includegraphics[width=3.2in, angle=0]{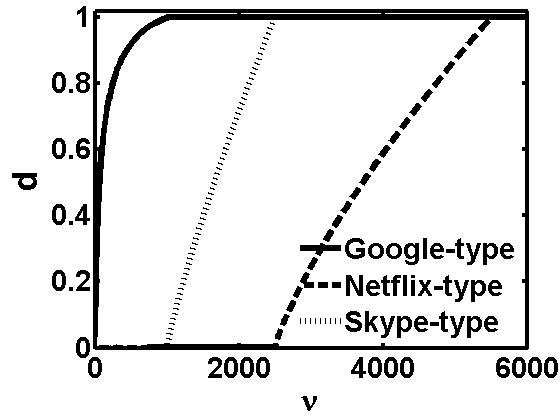}
\caption{Throughput under max-min fair mechanism.}
\label{figure:3CPs_maxmin}
\end{figure}

Figure \ref{figure:3CPs_maxmin} illustrates the rates and the corresponding demands of the three CPs under a max-min fair allocation mechanism. We vary the per capita capacity $\nu$ from $0$ to $6,000$.
We observe that when  $\nu$ increases from zero, the demand for Google-type content increases first, followed by the demand for Skype-type content and, the demand for Netflix-type content being the last.

\section{Monopolistic ISP Analysis} %
\label{sec:monopoly}
In this section, we start with the scenario where the last-mile capacity is controlled by a single monopolistic ISP  ${\mathit I}$. We analyze the ISP's strategy under which non-neutral service differentiation is allowed, and the corresponding best responses of the CPs. We derive the equilibria of the system and analyze the ISP's impact on the system congestion and the welfare of the consumers and the ISP itself.

\subsection{Non-Neutral Service Differentiation}\label{sec:differentiation}
We assume that the monopolistic last-mile ISP $\mathit I$ has a capacity of $\mu$.
This ISP can be a retail residential ISP, e.g. Comcast and Time Warner Cable, or a mobile operator, e.g. Verizon and AT\&T. Regardless of whether it is a wired or wireless provider, it serves as the last-mile service provider for the consumers. We assume that the ISP is allowed to allocate a fraction $\kappa \in [0,1]$ of its capacity to serve premium CPs and charge them at a rate $c \in [0,\infty)$ (dollar per unit traffic).
For a wired ISP, $\kappa$ can be interpreted as the percentage of capacity deployed for private peering points that charge a fee of $c$ per unit incoming traffic and $1-\kappa$ can be interpreted as the percentage of capacity deployed for public peering points where incoming traffic is charge-free. For a wireless ISP, $\kappa$ can be interpreted as the percentage of capacity devoted for the premium traffic that will be charged at a rate of $c$. The pair of parameters $(\kappa,c)$ can also be thought of a type of Paris Metro Pricing (PMP) \cite{odlyzko99paris,shetty10internet}, where one ordinary and another premium service class have capacities of $(1-\kappa) \mu$ and $\kappa \mu$ and charge $0$ and $c$ respectively.
In reality, content might be delegated via content distribution networks (CDNs), e.g. Akamai, or backbone ISPs, e.g. Level3  is a major tier-1 ISP that delivers Netflix traffic towards regional ISPs.
Therefore, in practice, the charge $c$ might be imposed on the delivering ISP, e.g. Level3, and then be recouped from the CP, e.g. Netflix, by its delivering ISP, e.g. Level3. Our model does not assume any form of the implementation.

We denote ${\cal O}$ and ${\cal P}$ as the two disjoint sets of CPs that join the ordinary and premium class respectively. We denote $v_i$ as CP $i$'s per unit traffic revenue. This revenue can be generated by advertising for media clients, e.g Google, or by selling products to online consumers, e.g. Amazon, or by providing services to consumers, e.g. Netflix and e-banking. Our model does not assume how the revenue is generated either. Each CP $i$'s utility function $u_i$ can be expressed as
\begin{equation}
 u_i(\lambda_i) =
 \begin{cases}
\displaystyle{v_i\lambda_i} & \text{if } i\in {\cal O},\\
\displaystyle{v_i\lambda_i - c\lambda_i} & \text{if } i\in {\cal P}.
\end{cases}
\label{equation:utility_function}
\end{equation}
We define $IS = c \lambda_{\cal P}$ as the ISP surplus (CP-side revenue)
and denote $\Psi$ as the per capita ISP surplus defined as
\[\Psi = \frac{IS}{M}  = \frac{c}{M} \lambda_{\cal P} = \frac{c}{M} \sum_{i\in {\cal P}} \lambda_i(\theta_i)= c \sum_{i\in {\cal P}} \alpha_i d_i(\theta_i)\theta_i.\]



\subsection{Content Provider's Best Response}
Given the ISP's decision $\kappa$ and $c$, each CP chooses whether to join the ordinary service class $\cal O$ or the premium class $\cal P$.
We denote $\rho_i$ as the per capita throughput over CP $i$'s user base, i.e. $\alpha_i M$ users, defined as
\begin{equation}
\rho_i(\nu,{\cal N}) = d_i(\theta_i(\nu,{\cal N})) \theta_i(\nu,{\cal N}).
\end{equation}

{\lemma Given a fixed set $\cal O$ of CPs in the ordinary class and a fixed set $\cal P$ of CPs in the premium class, a new CP $i$'s optimal strategy is to join the premium service class, if
\begin{equation}
(v_i-c) \ \rho_i\big(\kappa\nu,{\cal P}\cup \{i\}\big) \geq v_i  \ \rho_i\big((1-\kappa)\nu,{\cal O}\cup \{i\}\big).
\label{equation:optimal_decision}
\end{equation}
Moreover, when equality reaches, CP $i$  obtains the same utility in both service classes.
\label{lemma:optimal_decision}
}

\noindent {\bf Proof of Lemma \ref{lemma:optimal_decision}:}
When joining the ordinary service class, the throughput of CP $i$ is
\begin{IEEEeqnarray*}{c}
\lambda_i(M,(1-\kappa)\mu,{\cal O}\cup \{i\}) = \alpha_i M \rho_i(M,(1-\kappa)\mu,{\cal O}\cup \{i\}) \\
=  \alpha_i M \rho_i((1-\kappa)\nu,{\cal O}\cup \{i\}).
\end{IEEEeqnarray*}
When joining the premium service class, the throughput is
\begin{IEEEeqnarray*}{c}
\lambda_i(M,\kappa\mu,{\cal P}\cup \{i\}) = \alpha_i M \rho_i(M,\kappa\mu,{\cal P}\cup \{i\}) \\
=  \alpha_i M \rho_i(\kappa\nu,{\cal P}\cup \{i\}).
\end{IEEEeqnarray*}
Therefore, by Equation \ref{equation:utility_function}, CP $i$'s utility is
\begin{equation*}
 u_i =
 \begin{cases}
\displaystyle{v_i\alpha_i M \rho_i((1-\kappa)\nu,{\cal O}\cup \{i\})} & \text{if } i\in {\cal O},\\
\displaystyle{(v_i - c)\alpha_i M \rho_i(\kappa\nu,{\cal P}\cup \{i\})} & \text{if } i\in {\cal P}.
\end{cases}
\end{equation*}
By comparing the utilities that can be obtained in the two service classes, we obtain the condition (\ref{equation:optimal_decision}).
\done

\bigskip

Lemma \ref{lemma:optimal_decision} states that a CP will join the premium service class if that results higher profit, which is per-unit flow profit ($v_i-c$ for the premium class) multiplied by the per capita throughput $\rho_i$.
The above decision is clear for a CP only if all other CPs have already made their choices. To treat all CPs equally, we model the decisions of all CPs as a simultaneous-move game as part of a two-stage game.

\subsection{Two-Stage Strategic Game}
We model the strategic behavior of the ISP and the CPs as a two-stage game, denoted as a quadruple $(M,\mu, {\cal N}, {\mathit I})$.
\begin{enumerate}
\item {\it Players}: The ISP $\mathit I$ and the set of CPs $\cal N$.
\item {\it Strategies}: ISP $\mathit I$ chooses a strategy $s_{\mathit I}=(\kappa,c)$. Each CP $i$ chooses a binary strategy of whether to join the premium class. The CPs' strategy profile can be written as $s_{\cal N}=({\cal O},{\cal P})$, where ${\cal O} \cup {\cal P} = {\cal N}$ and  ${\cal O} \cap {\cal P} = \emptyset$.
\item {\it Rules}: In the first stage, ISP $\mathit I$ decides $s_{\mathit I} =(\kappa,c)$ and announces it to all the CPs. In the second stage, all the CPs make their binary decisions simultaneously and reach a joint decision  $s_{\cal N}=({\cal O},{\cal P})$.
\item {\it Outcome}: The set $\cal P$ of the CPs shares a capacity of $\kappa \mu$ and the set $\cal O$ of the CPs shares a capacity of $(1-\kappa)\mu$. Each CP $i\in \cal O$ gets a rate $\lambda_i(M,(1-\kappa)\mu,\cal O)$ and each CP $j\in \cal P$ gets a rate $\lambda_j(M,\kappa \mu,\cal P)$. 
\item {\it Payoffs}: Each CP $i$'s payoff is defined by the utility $u_i(\lambda_i)$ in Equation (\ref{equation:utility_function}). The ISP's payoff is the revenue $IS = c \lambda_{\cal P}$ 
    received from the premium class.
\end{enumerate}
If we regard the set of CPs as a single player that chooses a strategy $s_{\cal N}$, our two-stage game can be thought of a Stackelberg game \cite{osborne-course}. In this game, the first-mover ISP can take all the best-responses of the CPs into consideration and derive its optimal strategy $s_{\mathit I}$ using {\em backward induction} \cite{andreu-microeco}. Given any fixed strategy $s_{\mathit I}=(\kappa,c)$, the CPs derive their best strategies under a simultaneous-move game, denoted as $(M,\mu,{\cal N},s_{\mathit I})$. We denote $s_{\cal N} (M, \mu,{\cal N}, s_{\mathit I}) = ({\cal O},{\cal P})$ as a strategy profile of the CPs under the game $(M, \mu,{\cal N}, s_{\mathit I})$.
Technically speaking, when $\kappa=0$ or $1$, there is only one service class. When $\kappa=0$, we define the trivial strategy profile as $s_{\cal N}=({\cal N},\emptyset)$; when $\kappa=1$, although there is not a physical ordinary class, we define the trivial strategy profile as $s_{\cal N}=({\cal O},{\cal N}\backslash {\cal O})$, with ${\cal O} = \{i:v_i\leq c, i\in \cal N\}$ which defines the set of ISPs that cannot afford to join the premium class.
Based on Lemma \ref{lemma:optimal_decision}, we can define an equilibrium in the sense of a Nash or competitive equilibrium. To break a tie, we assume that a CP always chooses to join the ordinary service class when both classes provide the same utility.

{\definition A strategy profile $s_{\cal N}=({\cal O},{\cal P})$ is a Nash equilibrium of a game $(M,\mu,{\cal N},s_{\mathit I})$, if
\begin{equation}
\frac{v_i-c}{v_i}
 \begin{cases}
\displaystyle{\leq \frac{\rho_i\big((1-\kappa)\nu,{\cal O}\big)}{\rho_i\big(\kappa\nu,{\cal P}\cup \{i\}\big)} } & \text{if } i\in {\cal O}, \\
\\
\displaystyle{> \frac{\rho_i\big((1-\kappa)\nu,{\cal O}\cup \{i\}\big)}{\rho_i\big(\kappa\nu,{\cal P}\big)} } & \text{if } i\in {\cal P}.
\end{cases}
\label{equation:Nash_equilibrium}
\end{equation}
\label{definition:Nash_equilibrium}}

\subsection{Competitive Equilibrium}
Notice that a CP's joining decision to a service class might increase the congestion level and reduce the throughput of flows of that service class; however, if the number of CPs in a service class is big, an additional CP $i$'s effect will be marginal.
Analogous to the {\em pricing-taking assumption} \cite{andreu-microeco} in a competitive market, we can make a {\em throughput-taking assumption} for the CPs as follows.

{\assumption Any CP $i \notin \cal N$ makes an estimate $\tilde{\rho}_i(\nu,{\cal N})$ on its ex-post per capita throughput $\rho_i(\nu,{\cal N}\cup \{i\})$ in the decision-making under a competitive equilibrium.}

Based on the above throughput-taking assumption, we can define a competitive equilibrium of the CPs as follows.
{\definition A strategy profile $s_{\cal N}=({\cal O},{\cal P})$ is a competitive equilibrium of a game $(M,\mu,{\cal N},s_{\mathit I})$, 
 if 
\begin{equation}
\frac{v_i-c}{v_i}
 \begin{cases}
\displaystyle{\leq \frac{\rho_i\big((1-\kappa)\nu,{\cal O}\big)}{\tilde{\rho}_i\big(\kappa \nu,{\cal P}\big) }} & \text{if } i\in {\cal O}, \\
\\
\displaystyle{> \frac{\tilde{\rho}_i\big((1-\kappa)\nu,{\cal O}\big)}{\rho_i\big(\kappa\nu,{\cal P}\big)} } & \text{if } i\in {\cal P}.
\end{cases}
\label{inequality:optimal_competitive}
\end{equation}
\label{definition:competitive_equilibrium}}

The competitive equilibrium depends on how each CP $i$ calculates $\tilde{\rho}_i= d_i(\tilde{\theta}_i)\tilde{\theta}_i$, which boils down to an estimation of the ex-post throughput $\tilde{\theta}_i$.
This estimation depends on the rate allocation mechanism being used.
For example, under the max-min fair mechanism, CP $i$ can expect an achievable throughput of $\theta_{\cal N}=\max \{\theta_j:j\in \cal N\}$. Thus, CP $i$ can take this {\it throughput} as given and estimate that  $\tilde{\theta}_i = \min \{ \hat \theta_i, \theta_{\cal N} \}$.

To perform numerical evaluations, we will focus on competitive equilibria for two reasons. First, because the number of CPs in practice is big,  the congestion-taking assumption is valid. Second, the {\em common knowledge} assumption \cite{andreu-microeco} for reaching Nash equilibria might be questionable, because
CPs rarely know the characteristics of all other CPs in practice. Nevertheless, most of our results
 apply for both equilibrium definitions.
In the rest of the paper, unless we specifically indicate an equilibrium to be Nash or competitive, we use the term {\em equilibrium} to indicate both the Nash (Definition \ref{definition:Nash_equilibrium}) and the competitive equilibrium (Definition \ref{definition:competitive_equilibrium}).

{\theorem If $s_{\cal N}=(\cal O,\cal P)$ is an equilibrium of a game $(M,\mu,{\cal N},s_{\mathit I})$,  it is also a same type of equilibrium (Nash or competitive) of a game $(\xi M, \xi \mu,{\cal N},s_{\mathit I})$ for any $\xi>0$. \label{theorem:Nash_linear_scale}}

\noindent {\bf Proof of Theorem \ref{theorem:Nash_linear_scale}:}
Under the same strategy $s_{\mathit I} = (\kappa,c)$, the new ordinary class $(\xi M, \xi (1-\kappa)\mu, \cal O)$ and the new premium class $(\xi M, \xi \kappa \mu, \cal P)$ have the same per capita capacity $(1-\kappa)\nu$ and $\kappa \nu$ respectively as before. By Lemma \ref{lemma:theta}, the new system induces the same throughput $\theta_i$ as before. As a result, the solution $(\cal O,P)$ will induce the same values of $\rho_i$ and $\tilde \rho_i$, which is an estimate on $\rho_i$ based on $\theta_i$. Therefore, both sides of (\ref{equation:Nash_equilibrium}) and (\ref{inequality:optimal_competitive}) do not change and the equilibrium conditions still hold.
\done

\bigskip

Although a game $(M, \mu, {\cal N}, s_{\mathit I})$ might have multiple equilibria, we do not assume that it reaches a particular equilibrium. However, to make equilibria under the same per capita capacity $\nu$ consistent, we make the following assumption.
{\assumption If $s_{\cal N}=(\cal O,\cal P)$ is a realized equilibrium of a game $(M,\mu,{\cal N},s_{\mathit I})$, then it is also the realized equilibrium of the linearly scaled game $(\xi M, \xi \mu,{\cal N},s_{\mathit I})$ for any $\xi>0$.\label{assumption:equilibrium_consistency}}

The above assumption implies that when the ISP scales its capacity $\mu$ linearly and smoothly with its consumer size $M$, the CPs will not diverge abruptly into another equilibrium, if there exists any.
For the game $(M,\mu,{\cal N},s_{\mathit I})$ with strategy $s_{\mathit I}=(\kappa,c)$ and equilibrium $s_{\cal N}=(\cal O,P)$, the per capita consumer surplus $\Phi$ is a function of $\nu$, defined as
\[ \Phi(M,\mu,{\cal N},s_{\mathit I}) = \Phi(\nu,{\cal N},s_{\mathit I}) = \Phi((1-\kappa)\nu,{\cal O})+\Phi(\kappa\nu,{\cal P}).\]
Under the above assumption, the per capita consumer and ISP surplus will remain the same in linearly scaled games $\{(\xi M, \xi \mu, {\cal N}, s_{\mathit I}):\xi>0\}$ in equilibrium.

{\lemma Under Assumption \ref{assumption:equilibrium_consistency}, the per capita consumer surplus $\Phi$
satisfies
\begin{equation*}
\Phi(\nu, {\cal N},s_{\mathit I}) = \Phi(\xi M,\xi \mu,{\cal N},s_{\mathit I}), \quad \forall \ \xi>0.
\end{equation*}
\label{lemma:equal_per_capita_cs}
The above is true for the per capita ISP surplus $\Psi$ as well.}

\noindent {\bf Proof of Lemma \ref{lemma:equal_per_capita_cs}:}
Given $s_{\mathit I} = (\kappa,c)$ and $s_{\cal N}=(\cal O,P)$, the per capita consumer surplus $\Phi$ in system $(M,\mu, \cal N)$ is
\begin{equation*}
\Phi(M,\mu,{\cal N}, s_{\mathit I}) = \Phi(M,(1-\kappa)\mu,{\cal O}) + \Phi(M,\kappa\mu,{\cal P}).
\end{equation*}
By Theorem \ref{theorem:Phi_monotonicity}, we can rewrite the above as
\[\Phi(M,\mu,{\cal N}, s_{\mathit I}) = \Phi((1-\kappa)\nu,{\cal O}) + \Phi(\kappa\nu,{\cal P}).\]
By Assumption \ref{assumption:equilibrium_consistency}, we know that a linearly scaled system $(\xi M,\xi \mu,{\cal N})$ under the same strategy $s_{\mathit I}$ will induce the same equilibrium $s_{\cal N} = (\cal O,P)$. As a result, we have
\[\Phi(\xi M,\xi \mu,{\cal N}, s_{\mathit I}) = \Phi((1-\kappa)\nu,{\cal O}) + \Phi(\kappa\nu,{\cal P}) \quad \forall \xi>0.\]
Since the right hand side of the above equations are the same, we can express $\Phi$ as a function of $\nu$ as
\[\Phi(\nu,{\cal N}, s_{\mathit I}) = \Phi(\xi M,\xi \mu,{\cal N}, s_{\mathit I})  \quad \forall \xi>0.\]
Since $\Psi = c\sum_{i\in\cal P}\alpha_i \rho_i(\nu,\cal N)$, under a scaled system where $\nu$ and $\cal P$ do not change, $\Psi$ does not change either. \done

\bigskip

\subsection{Monopolistic ISP's Strategy}
In order to generate revenue, the ISP's optimal strategy would encourage more CPs to join its premium service class.


{\theorem In the game $(M,\mu,{\cal N},{\mathit I})$, for any $0\leq c<1$, strategy $s_{\mathit I}=(\kappa,c)$ is always dominated by $s^{1}_{\mathit I}=(1,c)$. If $\lambda_{{\cal P}} < \min \{\mu, \sum_{v_i \geq c} \hat\lambda_i\}$, $s_{\mathit I}$ is strictly dominated by $s^{1}_{\mathit I}$.
$s_{\mathit I}=(\kappa,c)$ is also dominated by $s'_{\mathit I}=(\kappa',c)$ with $\kappa'>\kappa$, if equilibrium $(\cal O',P')$ under $s'_{\mathit I}$ satisfies $\cal P\subseteq \cal P'$.
\label{theorem_kappa}}

\noindent {\bf Proof of Theorem \ref{theorem_kappa}:}
Under $s^1_{\mathit I} = (1,c)$, only the premium service class is provided and CPs will join the premium service class only if $v_i \geq c$. Therefore, the aggregate rate $\lambda^1_{\cal P} = \min \{\mu,\sum_{i\in {\cal P}_c}\hat\lambda_i\}$, where ${\cal P}_c = \{i:v_i\geq c\}$.
When keeping the same $c$, ${\cal P} \subseteq {\cal P}_c$ under any strategy $s_{\mathit I}$. We have
\[\lambda_{\cal P} = \min \{\kappa\mu,\sum_{i\in {\cal P}\subseteq {\cal P}_c}\hat\lambda_i\} \leq \min \{\mu,\sum_{i\in {\cal P}_c}\hat\lambda_i\} = \lambda^1_{\cal P}. \]
The above implies that the revenue $c\lambda_{\cal P} \leq c\lambda^1_{\cal P}$. Therefore, $s_{\mathit I}$ is dominated by $s^1_{\mathit I}$. If $\lambda_{\cal P} < \min \{\mu,\sum_{i\in {\cal P}_c}\hat\lambda_i\}$, $\lambda_{\cal P} < \lambda^1_{\cal P}$, and therefore $s_{\mathit I}$ is strictly dominated by $s^1_{\mathit I}$.

Similarly, for any $\kappa'>\kappa$ with ${\cal P}\subseteq {\cal P'}$, we have
\[\lambda_{\cal P} = \min \{\kappa\mu,\sum_{i\in {\cal P}\subseteq {\cal P'}}\hat\lambda_i\} \leq \min\{\kappa'\mu, \sum_{i\in\cal P'} \hat \lambda_i\} = \lambda_{\cal P'}.\]
Therefore, $s_{\mathit I}$ is also dominated by $s'_{\mathit I}$. \done

\bigskip

When the monopoly ISP increases $\kappa$, it improves the condition in the premium service class and in a new equilibrium, $\cal P'$ would attract more CPs to join than $\cal P$. Theorem \ref{theorem_kappa} states that the ISP would have incentives to increase $\kappa$ so as to maximize revenue.
The effect of increasing $\kappa$ is twofold: 1) more capacity is allocated to the premium class for sale, and 2) the reduced capacity in the ordinary class makes more ISPs switch to the premium class.
As a result, one of the optimal strategies of the monopolistic ISP is to always set $\kappa=1$. This implies that, if allowed, the selfish  ISP will only provide a charged service class $\cal P$ without contributing any capacity for the ordinary class $\cal O$.
Suppose the ISP is allowed to set $\kappa=1$, we first study its optimal price $c$ and its impact on
the consumer and ISP surplus.

We use the demand function of Equation (\ref{equation:demand_function}) and the max-min fair mechanism for our numerical simulations. 
We study a scenario of $1000$ CPs, whose $\alpha_i$, $\hat\theta_i$ and $v_i$ are uniformly distributed within $[0,1]$ and $\beta_i$ is uniformly distributed within $[0,10]$. To satisfy all unconstrained throughput for the CPs, the per capita capacity needs to be around $\nu = 250$.
Since throughput-sensitive applications, e.g. Skype, bring more utility to consumers in reality, we consider the consumer utility $\phi_i$ that is uniformly distributed within $[0, \beta_i]$ (the uniform distribution biases utility towards CPs with high throughput sensitivity while introducing some randomness)\footnote{We also used another setting of $\phi_i = U[0,U[0,10]]$, that has the same scale but is independent of $\beta_i$s for all the experiments in Section \ref{sec:monopoly} and \ref{sec:oligopoly}. All the results are similar with the current setting and are shown in the appendix.}.

\begin{figure}[!ht]
\centering
\includegraphics[width=2.3in, angle=0]{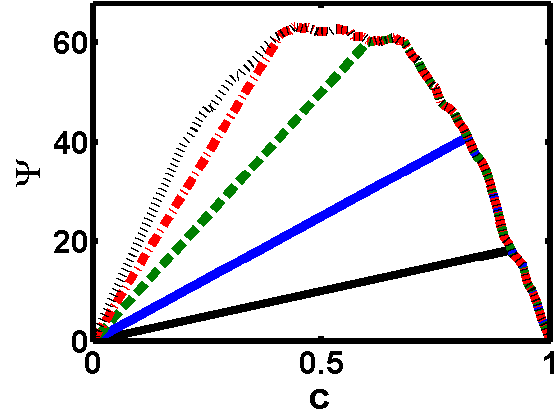}
\includegraphics[width=2.3in, angle=0]{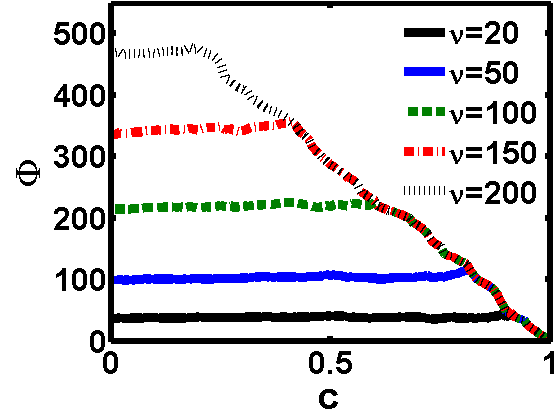}
\caption{Per capita surplus $\Psi$ and $\Phi$ under $\kappa=1$.}
\label{figure:Kappa2}
\end{figure}

\begin{figure*}[!ht]
\centering
\includegraphics[width=7.2in, angle=0]{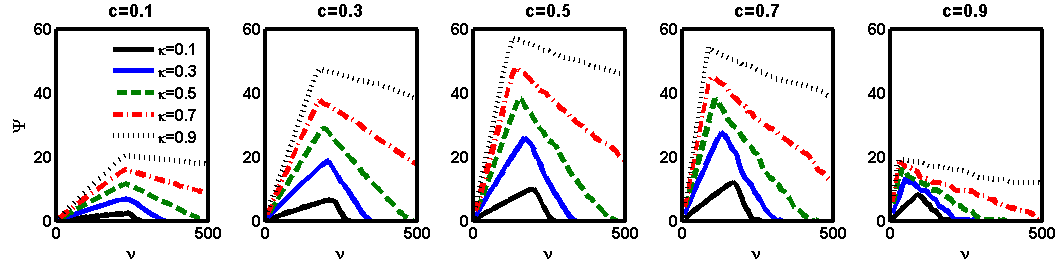}
\includegraphics[width=7.2in, angle=0]{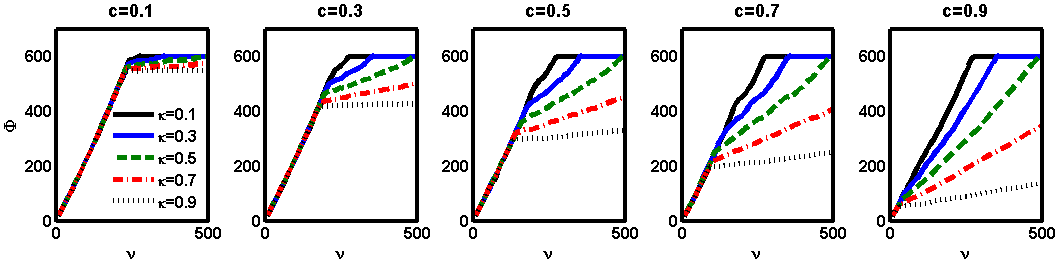}
\caption{Per capita surplus $\Psi$ and $\Phi$ under various strategies $s_{\mathit I}=(\kappa,c)$ versus per capita capacity $\nu$.}
\label{figure:monopoly}
\end{figure*}

Figure \ref{figure:Kappa2} plots $\Psi$ and $\Phi$ versus the ISP's pricing strategy $c$ under various per capita capacity $\nu$ ranging from $20$ to $200$. We observe three pricing regimes. \begin{enumerate}
\item When $c$ is small, $\Psi$ increases linearly, i.e. $\Psi = c\nu$. This happens when most of the CPs can afford to join the service and the entire capacity is fully utilized, i.e. $\lambda_{\cal P}=\mu$, resulting a high level of consumer surplus $\Phi$.
\item When $c$ is large, $\Psi$ drops sharply. This happens when only a small set of CPs can afford to join the service and the capacity is largely under-utilized, i.e. $\lambda_{\cal P}<\mu$, resulting a sharp drop in $\Phi$ accordingly.
\item When $\nu$ is abundant, e.g. $\nu=200$, there exists a pricing region where $\Psi$ increases sub-linearly and $\Phi$ decreases. Consequently, the ISP's optimal strategy ( $c\approx 0.45$) could intentionally keep more CPs away from the (only) service class and under-utilizes the capacity, which hurts the consumer surplus $\Phi$.
\end{enumerate}
As a result, the consumer surplus could be misaligned with a monopoly ISP's revenue when the capacity is abundant.


Figure \ref{figure:monopoly} illustrates $\Psi$ and $\Phi$ under various strategies $s_{\mathit I}=(\kappa,c)$ versus $\nu$ ranging up to $500$, which doubles the required capacity to serve all unconstrained throughput. For a fixed $c$ in each column, we identify three equilibrium regimes. \begin{enumerate}
\item When $\nu$ is small, $\Psi$ increases linearly and $\Phi$ increases accordingly. This happens when the premium class capacity is fully utilized, i.e. $\lambda_{\cal P} = \kappa \mu$.
\item When $\nu$ keeps increasing, $\Psi$ starts to decrease and $\Phi$ increases at a much slower rate. This happens when the premium class capacity is not fully utilized, i.e. $\lambda_{\cal P} < \kappa \mu$, and  more CPs move from $\cal P$ to $\cal O$.
\item When $\nu$ is large, $\Psi$ drops to zero for small values of $\kappa$, where $\Phi$ is maximized. This happens when $\cal P = \emptyset$ and $\cal O$'s capacity  is abundant enough to serve all CPs' unconstrained throughput. However, if $\kappa$ is big, e.g. $\kappa=0.9$, it guarantees some revenue for the ISP, but reduces the consumer surplus from its maximum.
\end{enumerate}
Further, under a fixed $\nu$, we observe that higher $\kappa$ induces higher revenue for the ISP (Theorem \ref{theorem_kappa}), even if that results an under-utilization of the premium class capacity and hurts the consumer surplus. When comparing different prices $c$, we observe that larger values of $c$ make the premium class becomes under-utilized faster, because fewer CPs can afford to join the premium class when necessary. However, when reaching the turning point where congestion starts to be relieved, 
$\kappa$ plays a major role, under which $\Phi$'s rate of increase depends on how much percentage $1-\kappa$ of capacity is allocated to the ordinary class $\cal O$.
Notice that only under the exceptional case where 1) the capacity is extremely scarce, and 2) CPs with high values of $v_i$s also have high values of $\phi_i$s for consumer utility, the ISP's optimal strategy of $\kappa=1$ and large value of $c$ might benefit the consumer surplus.

Under a fixed strategy $s_{\mathit I}$, $\Phi(\nu,{\cal N},s_{\mathit I})$ is not strictly non-decreasing in $\nu$ compared to the result of Theorem \ref{theorem:Phi_monotonicity}. The reason is that when $\nu$ varies, CPs might move between the service classes. In general, when $\nu$ changes a small amount such that $s_{\cal N} = (\cal O,P)$ does not change, the monotonicity still holds; however, when $\nu$ keeps increasing, CPs will move from $\cal P$ to $\cal O$, upon which $\Phi$ might drop at the spot. We characterize this discontinuity by the following metric.
\begin{equation}
\epsilon_{s_{\mathit I}} = \sup \{ \Phi(\nu_1,{\cal N},s_{\mathit I}) - \Phi(\nu_2,{\cal N},s_{\mathit I}):\nu_1 < \nu_2\}. \ \label{definition:epsilon}
\end{equation}
Notice that $\epsilon_{s_{\mathit I}}$ captures the largest vertical distance of a downward gap in the curve $\Phi(\nu,{\cal N},s_{\mathit I})$. From Figure \ref{figure:monopoly}, we observe that when $|\cal N|$ is large, $\epsilon_{s_{\mathit I}}$ is quite small, which indicates the general trend of increasing $\Phi$ with $\nu$.

{\bf Regulatory Implications:}
In the monopolistic scenario, the ISP with fixed capacity $\mu$ will maximize the percentage of capacity for a charged service class ($\kappa$) even if it is under-utilized. This implies that the ISP has the incentive to degrade service quality or avoid network upgrades or investments for the non-charged service class.
To remedy this problem, the network neutrality principle should be imposed to some extent to protect consumer surplus. In other words, the non-neutral service differentiation should be limited, especially when the capacity $\mu$ is not scarce, by two means: 1) limit the percentage of capacity devoted to a charged premium class\footnote{This was also suggested as a regulatory tool by an independent work of Shetty et. al. \cite{shetty10internet}.}, i.e. $\kappa$ cannot be too large, such that the CPs in the ordinary class can obtain an appropriate amount of capacity, and 2) limit the charge $c$ so that enough CPs would be able to join the premium class. The bottom line is that capacity under-utilization should be avoided under any case.

\section{Oligopolistic ISP Analysis}\label{sec:oligopoly}
In the previous section, we concentrate on a monopolistic ISP $\mathit I$ that has a capacity $\mu$ and uses a strategy $s_{\mathit I} = (\kappa,c)$. In this section, we extend our model to a set $\cal I$ of oligopolistic ISPs, each ${\mathit I} \in \cal I$ of which has a capacity $\mu_{\mathit I}$ and uses a strategy $s_{\mathit I} = (\kappa_{\mathit I},c_{\mathit I})$. We define $\mu = \sum_{\mathit I \in {\cal I}} \mu_{\mathit I}$ as the total system capacity, and $m_{\mathit I} = M_{\mathit I}/M$ and $\gamma_{\mathit I} = \mu_{\mathit I}/\mu$ as the market share and capacity share of ISP $\mathit I$. Our oligopolistic models have two major differences with the monopolistic model. First, since consumers connect to the Internet via one of the ISPs, they might make strategic decisions on which ISP to subscribe to. We denote $M_{\mathit I}$ as the consumer size  of each ISP $\mathit I\in \cal I$, where $\sum_{{\mathit I} \in \cal I} M_{\mathit I}$ equals the total consumer size $M$. Second, besides maximizing the premium service revenue from the CPs, a more important objective of any ISP $\mathit I\in\cal I$ is to maximize their market share $m_{\mathit I}$ of the consumers. This is what the last-mile ISPs can leverage on to generate the CP-side of the revenue in the first place.

Similar to the monopolistic ISP game $(M,\mu, {\cal N}, {\mathit I})$, we denote $(M,\mu, {\cal N}, {\cal I})$ as the two-stage oligopolistic ISP game, under which the set of ISPs $\cal I$ choose their strategies $s_{\cal I} = \{s_{\mathit I}:\mathit I\in\cal I\}$ simultaneously in the first stage, and then the set of CPs $\cal N$ and the $M$ consumers make their strategic decisions simultaneously in a second-stage game $(M,\mu,{\cal N},s_{\cal I})$. In the second-stage game, we denote $s_M = \{M_{\mathit I}:\mathit I\in\cal I\}$ as the consumers' strategy that determines all ISPs' market shares, and $s_{\cal N} = \{s_{\cal N}^{\mathit I} = ({\cal O}_{{\mathit I}},{\cal P}_{{\mathit I}}): \mathit I\in\cal I\}$ as the CPs' strategy, where each $s_{\cal N}^{\mathit I}$ denotes the decision made at ISP $\mathit I$.

We denote $\Phi_{\mathit I}$ as the per capita consumer surplus achieved at ISP $\mathit I$, defined as
$\Phi_{\mathit I}(M_{\mathit I},\mu_{\mathit I},{\cal N},s_{\mathit I}) = \Phi_{\mathit I}(\nu_{\mathit I},{\cal N},s_{\mathit I}) = \Phi((1-\kappa_{\mathit I})\nu_{\mathit I},{\cal O}_{\mathit I}) + \Phi(\kappa_{\mathit I}\nu_{\mathit I},{\cal P}_{\mathit I})$, where $\nu_{\mathit I} = \mu_{\mathit I} / M_{\mathit I}$.
We assume that consumers will move towards the ISPs that provide higher per capita surplus as follows.

{\assumption Under any fixed strategy profile $s_{\cal I}$ and $s_{\cal N}$,
for any pair of ISPs $\mathit I,J\in \cal I$, consumers will move from  $\mathit I$ to  $\mathit J$ if $\Phi_{\mathit I}<\Phi_{\mathit J}$. This process stops when $\Phi_{\mathit I} = \Phi_{\cal I} \ \forall \mathit I\in\cal I$ for some system-wise per capita consumer surplus $\Phi_{\cal I}$.
\label{assumption:consumer_mobility}}

Although consumers might not be totally elastic or/and accessible to all available ISPs in practice, our assumption takes a macro perspective and assumes that if an ISP provides worse user-experience on average, there must exist some consumers who can and will move to other better ISPs.
Based on Assumption \ref{assumption:consumer_mobility}, we define the equilibrium of the second-stage game $(M,\mu,{\cal N},s_{\cal I})$ as follows. {\definition A strategy profile $(s_M,s_{\cal N})$ is an equilibrium of the multi-ISP game $(M,\mu,{\cal N},s_{\cal I})$ if 1) for any $\mathit I\in\cal I$, $s_{\cal N}^{\mathit I}$ is an equilibrium of the single-ISP game $(M_{\mathit I},\mu_{\mathit I},{\cal N},s_{\mathit I})$, and 2) $\Phi_{\mathit I}= \Phi_{\mathit J}$ for any $\mathit I,J \in\cal I$.
\label{definition:second-stage_equilibrium}}

\subsection{Duopolistic ISP Game}\label{sec:public-option}\label{subsec:duopoly}
We first study a two-ISP game with ${\cal I} = \{\mathit I, J\}$. Before that, we formally define a Public Option ISP as follows.
{\definition A Public Option ISP $PO$ is an ISP that uses a fixed strategy $s_{\mathit PO} = (0,0) $ and does not divide its capacity or charge the CPs.
\label{definition:public-option}}

\begin{figure}[ht]
\centering
\includegraphics[width=4.2in, angle=0]{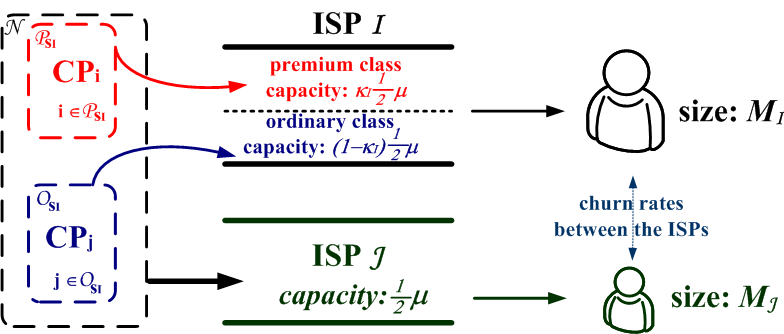}
\caption{A two-ISPs model.}
\label{figure:competition}
\end{figure}
We assume that ISP $\mathit J$ is a Public Option ISP.
Figure \ref{figure:competition} illustrates an example of the above duopolistic model, where both ISPs have the same amount of capacity, the CPs choose a service class at ISP $\mathit I$ and the consumers move between the ISPs.
The above setting of the duopolistic game applies for two real scenarios. First, it models the competition between two ISPs, where one of them is actively a Public Option ISP and the other actively manages a non-neutral service differentiation.
Second, it also models a situation where a single ISP owns the entire last-mile capacity $\mu$; however, by regulation~\cite{unbundling-wikipedia}, it is required to lease its capacity to other service providers that do not own the physical line.  The leasing ISP might be technologically limited from providing service differentiation on the leased capacity, but actually have customers in the region.
For both scenarios, we will answer 1) whether the non-neutral ISP could obtain substantial advantages over the neutral Public Option ISP (or whether the Public Option could survive under competition), and 2) how the competition is going to impact the consumer surplus.

We study the same set of $1000$ CPs as in the previous section.
We further assume that $\mu_{\mathit I}=\mu_{\mathit J}=\mu/2$ in our numerical example. We take the same route to numerically evaluate the competitive equilibria of the CPs under $\kappa_{\mathit I}=1$.

\begin{figure}[!ht]
\centering
\includegraphics[width=2.3in, angle=0]{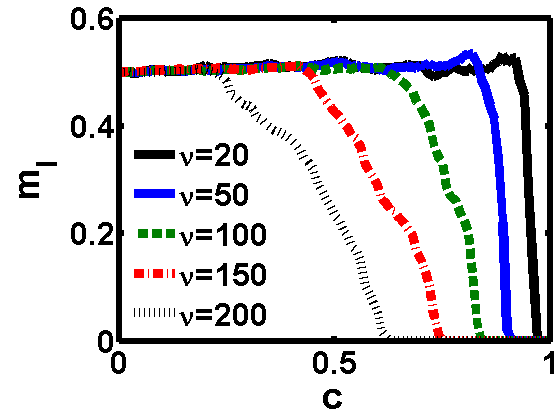}
\includegraphics[width=2.3in, angle=0]{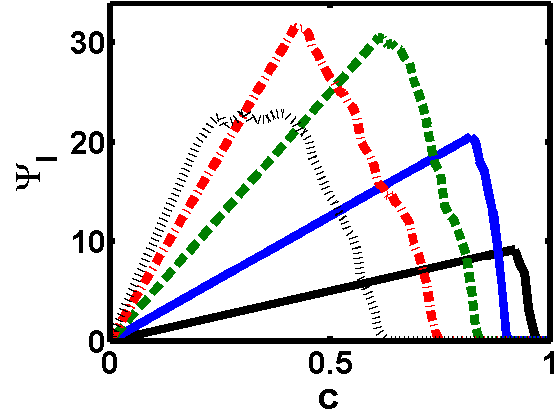}
\includegraphics[width=2.3in, angle=0]{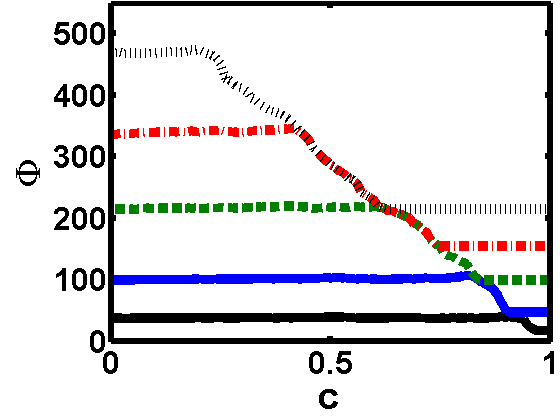}
\caption{ISP $\mathit I$'s market share $m_{\mathit I}$ and per capita surplus $\Psi_{\mathit I}$ and
per capita consumer surplus $\Phi$ under $\kappa=1$.}
\label{figure:Kappa_Oli}
\end{figure}

\begin{figure*}[!ht]
\centering
\includegraphics[width=7.2in, angle=0]{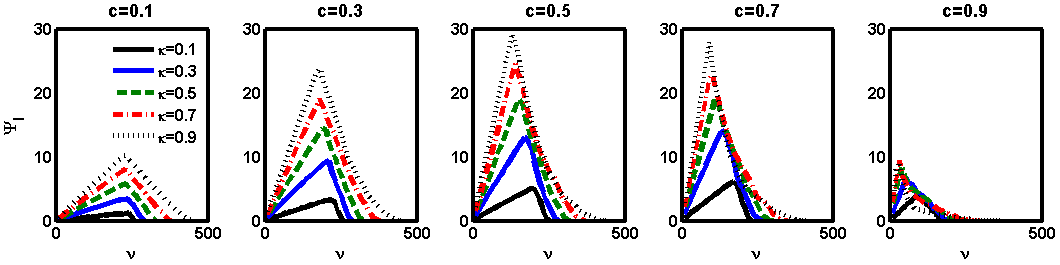}
\includegraphics[width=7.2in, angle=0]{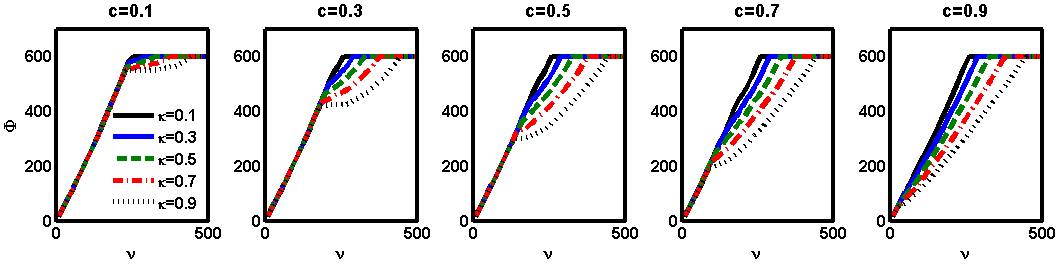}
\includegraphics[width=7.2in, angle=0]{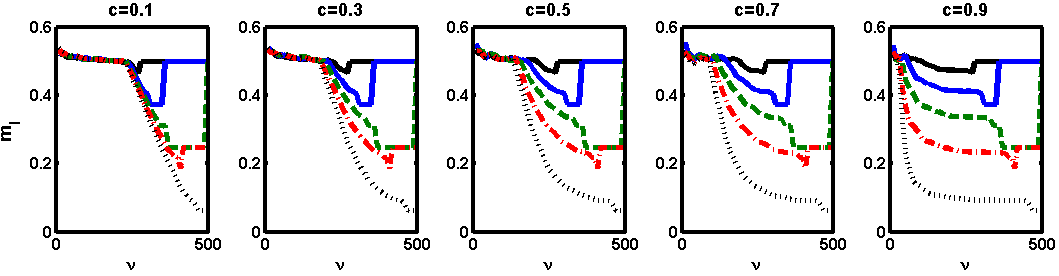}
\caption{Per capita surplus $\Psi$, $\Phi$ and market share $m_{\mathit I}$ under various strategies $s_{\mathit I}=(\kappa,c)$ vs. per capita capacity $\nu$.}
\label{figure:oligopoly}
\end{figure*}

Figure \ref{figure:Kappa_Oli} plots ISP $\mathit I$'s market share $m_{\mathit I}$, per capita surplus $\Psi_{\mathit I}$, defined as $\Psi_{\mathit I} = \frac{c}{M} \lambda_{{\cal P}_{\mathit I}}$, and $\Phi$ versus ISP $\mathit I$'s charge $c_{\mathit I}$. By the same reasons as before, the revenue of ${\mathit I}$ increases linearly when its capacity is fully utilized, i.e. $\lambda_{\cal P_{\mathit I}} = \kappa_{\mathit I} \mu_{\mathit I}$. However, we observe three differences: 1) after $\lambda_{\cal P_{\mathit I}}$ drops below $\kappa_{\mathit I} \mu_{\mathit I}$, $\Psi_{\mathit I}$ drops to zero much steeper than before, 2) $\Phi_{\mathit I}$ does not drop down to zero when $c_{\mathit I}$ increases to $1$, and 3) the maximum $\Psi_{\mathit I}$ is lower in the case of $\nu = 200$ than in the case of $\nu = 150$, which means that under $\kappa_{\mathit I}=1$, capacity expansion could reduce ISP $\mathit I$'s revenue from the CPs. All these observations can be explained by checking at the market share of ISP $\mathit I$ in the left sub-figure. The market share $m_{\mathit I}$ starts to increase with $c_{\mathit I}$ until ISP $\mathit I$'s capacity becomes under-utilized, i.e. $\lambda_{\cal P_{\mathit I}} < \kappa_{\mathit I} \mu_{\mathit I}$.
Afterwards, the market share drops dramatically. This explains that under congestion, i.e. $\lambda_{\cal P_{\mathit I}} = \kappa_{\mathit I} \mu_{\mathit I}$, by increasing $c_{\mathit I}$, ISP $\mathit I$ restricts the number of CPs in its service class and maintains less congestion, which could result higher consumer surplus, and therefore, attract more consumers from ISP $\mathit J$. After $\lambda_{\cal P_{\mathit I}}$ drops below $\kappa_{\mathit I} \mu_{\mathit I}$, further increase of $c_{\mathit I}$ reduces the number of CPs in the service as well as the total throughput. This reduces consumer surplus, and therefore, consumers start to depart from ISP $\mathit I$ to $\mathit J$. When $c_{\mathit I}$ reaches $1$, no CP survives in $\mathit I$'s service class and all consumers move to ISP $\mathit J$, which guarantees a non-zero consumer surplus in equilibrium.

Parallel to Figure \ref{figure:monopoly}, Figure \ref{figure:oligopoly} illustrates the per capita surplus $\Psi_{\mathit I}$, $\Phi$ and ISP $\mathit I$'s market share $m_{\mathit I}$ under various strategies $s_{\mathit I}$ versus $\nu$ ranging up to $500$. Compared to the monopolistic case, we observe two differences in $\Psi_{\mathit I}$ and $\Phi$: 1) under any strategy $s_{\mathit I}$, ISP $\mathit I$'s revenue drops sharply to zero after reaching a maximum point where $\lambda_{\cal P_{\mathit I}}$ drops below $\kappa_{\mathit I} \mu_{\mathit I}$, and 2) the increase of consumer surplus does not get affected by ISP $I$'s strategy too much.
By observing the market share of ISP $\mathit I$, we identify two capacity regimes. First, when $\nu$ is extremely scarce, the differential pricing slightly benefits the consumer; and therefore, ISP $\mathit I$ can obtain a slightly larger percentage of the market \footnote{Notice that this is because by limiting the number of CPs in the premium class, the proportion of throughput-sensitive traffic is larger, which gives higher utility to consumers.
}.
Second, when the per capita capacity $\nu$ is abundant, ISP $\mathit I$ obtains at most an equal share of the market if it uses a small value of $\kappa$. Under this case, the capacity under $\cal O$ can support half of the population's unconstrained throughput and in fact, the premium class is empty, i.e. $\cal P = \emptyset$. As a result, ISP $\mathit I$ follows the Public Option ISP by using some kind of neutral policy (small $\kappa$) and maximizes the consumer surplus.



{\theorem In the duopolistic game $(M,\mu,\cal N, I)$, where an ISP $\mathit J$ is a Public Option, i.e. $s_{\mathit J}=(0,0)$, if $s_{\mathit I}$ maximizes $M_{\mathit I}$ under an equilibrium $(s_M,s_{\cal N})$, then the per capita consumer surplus $\Phi_{\cal I}$ is also maximized under that equilibrium.
\label{theorem:optimality_two_ISPs}}

\noindent {\bf Proof of Theorem \ref{theorem:optimality_two_ISPs}:} 
For any strategy $s'_{\mathit I} \neq s_{\mathit I}$, we have two cases to analyze. First, if $M'_{\mathit I}=M_{\mathit I}$, then $M'_{\mathit J}=M - M'_{\mathit I}=M - M_{\mathit I} = M_{\mathit J}$.
Given the same market share for the public option ISP, it induces the same per capita consumer surplus $\Phi'_{\mathit J} = \Phi_{\mathit J}$. In equilibrium, we have $\Phi'_{\cal I}=\Phi'_{\mathit J} = \Phi_{\mathit J} = \Phi_{\cal I}$.
Second, if $M'_{\mathit I}<M_{\mathit I}$, then $M'_{\mathit J}=M - M'_{\mathit I} > M - M_{\mathit I}=M_{\mathit J}$. Given a larger market share for the public option ISP, the per capita capacity reduces, i.e. $\nu'_J<\nu_J$. By Theorem \ref{theorem:Phi_monotonicity}, the public option will not induce a larger per capita consumer surplus, i.e. $\Phi'_{\mathit J} \leq \Phi_{\mathit J}$. Thus, we have $\Phi'_{\cal I}=\Phi'_{\mathit J} \leq \Phi_{\mathit J} = \Phi_{\cal I}$ in equilibrium.
\done

\bigskip

Theorem \ref{theorem:optimality_two_ISPs} implies that the existence of a Public Option ISP is superior to a network neutral situation, where $s_{\mathit I}=(0,0)$. This is because given the freedom of choosing an optimal $s_{\mathit I}$ to maximize market share, ISP $\mathit I$'s strategy will induce a maximum consumer surplus under $s_{\mathit J}=(0,0)$.

Based on our results, we answer the previously raised two questions: 1) The non-neutral ISP cannot win substantially over the Public Option ISP, which can still be profitable under the competition, confirming the independent findings from \cite{dovrolis08can}. 2) Regardless of the capacity size, the competition induces higher consumer surplus in equilibrium than under network neutral regulations.
The strategic ISP could obtain slightly over $50\%$ of the market; however, if it differentiates services in the way that hurts consumer surplus, its market share will drop sharply.

{\bf Regulatory Implications:} In the duopolistic scenario with one of the ISPs being a Public Option, contrary to the monopolistic case, the non-neutral strategy $s_{\mathit I}$ is always aligned with the consumer surplus (Theorem \ref{theorem:optimality_two_ISPs}). This result shows an interesting alternative to remedy the network neutrality issue under a monopolistic market. Instead of enforcing the ISP to follow network neutrality, the government (or a private organization, since we independently verify it can be profitable) can provide the consumers with a Public Option ISP that is neutral to all CPs. Given such a neutral entity in the market, consumers will move to their public option if it provides higher consumer surplus than the non-neutral ISP that uses differential pricing to the CPs. Meanwhile, in order to maximize its market share, the non-neutral ISP will adapt its strategy to maximize consumer surplus. In conclusion, the introduction of a Public Option ISP is superior to network neutral regulations under a monopolistic market, since its existence aligns the non-neutral ISP's selfish interest with the consumer surplus.



\subsection{Oligopolistic ISP Competition Game}
After analyzing the duopolistic game between a non-neutral and a Public Option ISP, we further consider a deregulated market under which all ISPs make non-neutral strategies.
We consider a multi-ISP game under which each ISP ${\mathit I}$ chooses a strategy $s_{\mathit I}$ to maximize its market share $m_{\mathit I}$.

We first consider a homogenous strategy $s=(\kappa,c)$, which can be a preferred or regulated strategy, used by all ISPs.

{\lemma If $s_{\cal I}=\{s_{\mathit I} = s : \mathit I\in \cal I\}$ for some strategy $s=(\kappa,c)$, then $\{m_{\mathit I} = \gamma_{\mathit I}, s_{\cal N}^{\mathit I} = s_{\cal N}(M,\mu,{\cal N},s):\mathit I \in \cal I \}$ is an equilibrium of the game $(M,\nu,{\cal N},s)$.\label{lemma:market_share}}

Lemma \ref{lemma:market_share} shows a symmetric equilibrium where market share $m_{\mathit I}$ is proportional to  capacity $\mu_{\mathit I}$. It implies that ISPs will have incentives to invest and expend capacity so as to obtain a larger market share. This equilibrium could be reached when ISPs simply mimic one another's strategy. A further question is whether the competition of market share among the ISPs would induce equilibria where consumer surplus is high. We denote $s_{\mathit -I}$ as the strategy profile of the ISPs other than ISP $\mathit I$. Similar to the definition of $\epsilon_{s_{\mathit I}}$ in Equation (\ref{definition:epsilon}), we define $\delta_{s_{\mathit I}} = \sup \{ m_1 - m_2: \Phi(\nu_1,{\cal N},s_{\mathit I}) \leq \Phi(\nu_2,{\cal N},s_{\mathit I}) \}$ and $\epsilon_{s_{\mathit -I}} = \max \{\epsilon_{s_{\mathit J}}:{\mathit J} \in {\cal I} \backslash \{\mathit I\} \}$.

\noindent {\bf Proof of Lemma \ref{lemma:market_share}:}
When $M_{\mathit I} = \gamma_{\mathit I} M$ and $s_{\mathit I} = s$ for all $\mathit I \in\cal I$, the single-ISP game $(M_{\mathit I},\mu_{\mathit I},{\cal N},s)$ is a linearly scaled game of $(M,\mu,{\cal N},s)$. By Theorem \ref{theorem:Nash_linear_scale}, we know that
\[s^{\mathit I}_{\cal N} = s_{\cal N}(M,\mu,{\cal N},s) = s_{\cal N}(M_{\mathit I},\mu_{\mathit I},{\cal N},s_{\mathit I}), \quad \forall \ \mathit I\in\cal I.\]
By Lemma \ref{lemma:equal_per_capita_cs}, we know that
\[\Phi_{\mathit I} = \Phi(M_{\mathit I},\mu_{\mathit I},{\cal N},s_{\mathit I}) = \Phi(M,\mu,{\cal N},s), \quad \forall \ \mathit I\in\cal I.\]
The above satisfies the two conditions of an equilibrium in Definition \ref{definition:second-stage_equilibrium} and concludes the proof.
\done

\bigskip

{\theorem Under any fixed strategy profile $s_{\mathit -I}$, if $\mathit I$'s strategy $s_{\mathit I}$ is a best-response to $s_{\mathit -I}$ that maximizes its market share $m_{\mathit I}$ in the game $(M,\mu,{\cal N},s_{\cal I})$,
then $s_{\mathit I}$ is a $\epsilon_{s_{\mathit -I}}$-best-response for the per capita consumer surplus $\Phi_{\cal I}$, i.e.
\[ \Phi_{\cal I} \geq \Phi'_{\cal I} - \epsilon_{s_{\mathit -I}}, \quad \forall s'_{\mathit I} \neq s_{\mathit I}. \]
Moreover, if $s_{\mathit I}$ is a best-response that maximizes consumer surplus $\Phi_{\cal I}$ in the game $(M,\mu,{\cal N},s_{\cal I})$, then $s_{\mathit I}$ is a $\delta_{s_{\mathit I}}$-best-response for the market share $m_{\mathit I}$, i.e.
\[ m_{\mathit I} \geq m'_{\mathit I} - \delta_{s_{\mathit I}},  \quad \forall s'_{\mathit I} \neq s_{\mathit I}. \]
\label{theorem:multiISP_best_response}}

\noindent {\bf Proof of Theorem \ref{theorem:multiISP_best_response}:}
If $s_{\mathit I}$ maximizes the market share $M_{\mathit I}$, for any strategy $s'_{\mathit I}\neq s_{\mathit I}$, we have $M'_{\mathit I}\leq M_{\mathit I}$. Therefore, there exist an ISP $\mathit J \neq I$ such that $M'_{\mathit J}\geq M_{\mathit J}$. This implies $\nu'_{\mathit J}\leq \nu_{\mathit J}$. By the definition of $\epsilon_{s_{\mathit J}}$ in Equation (\ref{definition:epsilon}), we have $\Phi'_{\mathit J} -\Phi_{\mathit J} \leq \epsilon_{s_{\mathit J}}$, or equivalently $ \Phi_{\mathit J} \geq \Phi'_{\mathit J} - \epsilon_{s_{\mathit J}}$. Thus,
\[ \Phi_{\cal I} = \Phi_{\mathit J} \geq \Phi'_{\mathit J} - \epsilon_{s_{\mathit J}} = \Phi'_{\cal I} - \epsilon_{s_{\mathit J}} \geq \Phi'_{\cal I} - \epsilon_{s_{\mathit -I}}. \]

If $s_{\mathit I}$ maximizes the consumer surplus $\Phi_{\cal I}$, for any strategy $s'_{\mathit I}\neq s_{\mathit I}$, we have $\Phi'_{\cal I}\leq \Phi_{\cal I}$. By the definition of $\delta_{s_{\mathit I}}$, we have $m'_{\mathit I} - m_{\mathit I} \leq \delta_{s_{\mathit I}}$, or equivalently $ m_{\mathit I} \geq  m'_{\mathit I} - \delta_{s_{\mathit I}}$.
\done

\bigskip

Theorem \ref{theorem:multiISP_best_response} states that, given the fixed strategies of all other ISPs, an ISP's best-responses to maximize its market share and to maximize the consumer surplus are closely aligned.
Parallel to Theorem \ref{theorem:optimality_two_ISPs}, it shows that an ISP's selfish interest is, although not perfectly, aligned with the consumer surplus under competition. Technically, the $\epsilon_{s_{\mathit -I}}$ imperfection is due to the discontinuity of $\Phi(\nu,{\cal N},s_{\mathit I})$ in $\nu$. When $\epsilon_{s_{\mathit -I}}$ approaches zero, $\Phi$ will be non-decreasing mostly and the objectives of maximizing market share and maximizing consumer surplus will converge.

{\definition A strategy profile $s_{\cal I} = \{s_{\mathit I}:\mathit I\in\cal I\}$ is a {\em market share} Nash equilibrium of the game $(M,\mu,s_{\cal N},{\cal I})$ if for any $\mathit I\in\cal I$ and any strategy $s'_{\mathit I}\neq s_{\mathit I}$, the market share $m_{\mathit I}$ satisfies
$m_{\mathit I}(s'_{\mathit I},s_{\mathit -I}) \leq m_{\mathit I}(s_{\mathit I},s_{\mathit -I})$.
Similarly, $s_{\cal I}$ is a {consumer surplus} Nash equilibrium of the game $(M,\mu,s_{\cal N},{\cal I})$ if for any $\mathit I\in\cal I$ and any strategy $s'_{\mathit I}\neq s_{\mathit I}$, the consumer surplus $\Phi_{\cal I}$ satisfies $\Phi_{\cal I}(s'_{\mathit I},s_{\mathit -I}) \leq \Phi_{\cal I}(s_{\mathit I},s_{\mathit -I})$.
}

{\corollary If $s_{\cal I}$ is a market shares Nash equilibrium of the oligopolistic game $(M,\mu,{\cal N,I})$, then it is also a consumer surplus $\epsilon_{s_{\cal I}}$-Nash equilibrium, where $\epsilon_{s_{\cal I}} = \max \{\epsilon_{s_{\mathit I}} :\mathit I\in\cal I\}$. Conversely, if $s_{\cal I}$ is a consumer surplus Nash equilibrium, then it is also a market share $\delta_{s_{\cal I}}$-Nash equilibrium, where $\delta_{s_{\cal I}} = \max \{\delta_{s_{\mathit I}} :\mathit I\in\cal I\}$.
\label{corollary:epsilon_Nash}}

\noindent {\bf Proof of Corollary \ref{corollary:epsilon_Nash}:}
If $s_{\cal I}$ is a market share Nash equilibrium, by definition, each $s_{\mathit I}$ is a market share best response of $s_{\mathit -I}$. Therefore, by Theorem \ref{theorem:multiISP_best_response}, we have
\[ \Phi_{\cal I} \geq \Phi'_{\cal I} - \epsilon_{s_{\mathit -I}} \geq \Phi'_{\cal I} - \epsilon_{s_{\cal I}}, \quad \forall s'_{\mathit I} \neq s_{\mathit I}, \]
which concludes that $s_{\cal I}$ is a consumer surplus $\epsilon_{s_{\cal I}}$-Nash equilibrium.
If $s_{\cal I}$ is a consumer surplus Nash equilibrium, by definition, each $s_{\mathit I}$ is a consumer surplus best response of $s_{\mathit -I}$. Therefore, by Theorem \ref{theorem:multiISP_best_response}, we have
\[ m_{\mathit I} \geq m'_{\mathit I} - \delta_{s_{\mathit I}} \geq m'_{\mathit I} - \delta_{s_{\cal I}},  \quad \forall s'_{\mathit I} \neq s_{\mathit I}, \]
which concludes that $s_{\cal I}$ is a market share $\delta_{s_{\cal I}}$-Nash equilibrium.
\done

\bigskip

As a direct consequence of Theorem \ref{theorem:multiISP_best_response}, Corollary \ref{corollary:epsilon_Nash} addresses that the objectives of maximizing market share and maximizing consumer surplus are also closely aligned under Nash equilibria of the oligopolistic game $(M,\mu,{\cal N,I})$.

{\bf Regulatory Implications:} In the oligopolistic scenario, all ISPs' optimal strategies are closely aligned with the consumer surplus. Even some ISPs use sub-optimal decisions, any remaining ISPs' optimal strategy would still nearly maximize the system consumer surplus (Theorem \ref{theorem:multiISP_best_response}).
This alignment with consumer surplus also sustains under Nash equilibria of the multi-ISP competition game (Corollary \ref{corollary:epsilon_Nash}).
In conclusion, network neutral regulations are not needed or should not be imposed under a competition market; otherwise, the achieved consumer surplus would be sub-optimal compared to what can be achieved in the efficient Nash equilibria. However, regulations should enforce the ISPs to be transparent in the sense that each ISP's capacity and strategy should be common knowledge to all ISPs. This would help the market converge to an efficient equilibrium more easily.

\section{Related Work}
Despite of its short history, plenty of work on network neutrality can be found in computer science \cite{jon07net,musacchio07network,altman10application,dovrolis08can,shetty10internet,ma10internet}, economics \cite{choi10net,katz07economics}, and law \cite{tim05nn,sidak06consumer} literature.

From an economics perspective, Sidak \cite{sidak06consumer} looked at the network neutrality regulation from consumer welfare's point of view and argued that differential pricing is essential to the maximization of welfare. We also focus on the consumer welfare and seek the conditions under which ISPs' strategy would be aligned with consumer welfare. Choi et al. \cite{choi10net} analyzed the effect of neutral regulations on ISPs' investment incentive and found that capacity expansion decreases the sale price of the premium service. This coincides with our finding under the monopolistic scenario; however, under oligopolistic competitions, we find that ISPs do have incentives to increase capacity so as to maximize market share.

From an engineering perspective, Dhamdhere et al. \cite{dovrolis08can} took a profitability perspective and concluded that the ISPs can still be profitable without violating network neutrality. This supports our proposal of a Public Option ISP that can be implemented and sustained by either a government or a private organization.
Crowcroft \cite{jon07net} reviewed the technical aspects of network neutrality and concluded that ``perfect'' network neutrality has never been and should not be engineered. We share the same view that under competition, network neutrality regulation is not necessary; while, under a monopolistic market, a non-regulatory alternative can be a Public Option ISP that incentivizes the existing ISP to maximize consumer surplus.

From a modeling point of view, Musacchio et al. \cite{musacchio07network} considered advertising CPs and also used a two-stage model under which ISPs move first. Their focus was primarily on a monopolistic ISP.
Caron et al. \cite{altman10application} modeled differentiated pricing for two application type. Our model captures more applications types with parameters of popularity ($\alpha_i$), maximum throughput ($\hat\theta_i$) and sensitivity to the throughput ($\beta_i$).  Another departure in our approach from previous analyses is the way we model traffic and congestion in the network. Traditionally, the classical $M/M/1$ formula for delay has been used to abstract out traffic and congestion \cite{choi10net} in economic analyses. Our view is that a more appropriate approach is to more faithfully model closed loop protocols like TCP that carry most of the traffic on the Internet. Shetty et al. \cite{shetty10internet} used a similar PMP-like two-class service differentiations and considered capacity planning, regulation as well as differentiated pricing to consumers. Our differentiated pricing focuses on the CP-side, where the CPs choose service classes and consumers choose ISPs.

From a regulatory aspect, Wu \cite{tim05nn} surveyed the discriminatory practices, e.g. selectively dropping packets, of broadband and cable operators and proposed solutions to manage bandwidth and police ISPs so as to avoid discrimination. Shetty et al. \cite{shetty10internet} proposed a simple regulatory tool to restrict the percentage of capacity the ISPs dedicate to a premium service class.  Economides et al. \cite{economides12network} compared various regulations for quality of service, price discrimination and exclusive contracts, and drew conclusions on desirable regulation regimes.  Ma et al. \cite{ma10internet} considered the ISP settlement aspect and advocated the use of Shapley value as profit-sharing mechanism to encourage ISPs to maximize social welfare. Our proposal of a Public Option ISP, on the other hand, is an non-regulatory alternative to the network neutral regulations.


\section{Discussion and Conclusions}\label{sec:conclusions}
In a monopolistic market, the ISP's selfish non-neutral strategy hurts
consumer surplus. Although network neutral regulation can improve
consumer surplus under that case, we find a better non-regulatory
alternative which is to introduce a Public Option ISP. The existence
of a Public Option ISP incentivizes the existing ISP's strategy to be
aligned with consumer surplus, and achieve higher consumer surplus
than that under network neutral regulations. In an oligopolistic
competition, market force influences ISPs' non-neutral strategies to
be aligned with consumer surplus
and ISPs will get market shares proportional to their capacities.
Under this case, the existence of a Public Option ISP would be
sub-optimal compared to the efficient Nash equilibria;
however, its damage is very limited because the Public
Option ISP would be the only one that uses a sub-optimal strategy,
where all other ISPs can adapt to optimal strategies and more
consumers will move from the Public Option to better and non-neutral
ISPs.
Of course, there is no reason why the Public Option ISP
cannot perform the price discrimination that aligns with the consumer surplus, which induces
an efficient Nash equilibrium in theory. However, implementing a neutral public option
will avoid mistakes or accidental ``collusion'' with the existing ISPs in the market.
In contrast, if network neutral regulation is enforced, all ISPs
will have to perform an neutral but inefficient strategy, which could
reduce the consumer surplus substantially.

Theoretically speaking, the existence of a Public Option ISP will be
effective if $\mu_{PO}>0$, regardless how large its capacity is. This
is because, in the idealized game model, we assume that an ISP's sole
objective is to maximize its market share. In practice, ISPs will
trade off its market share with potential revenue from the CPs, which
depends on the characteristics of the CPs, e.g. their profit margin
and throughput sensitivity, and the condition of the system, e.g. the
available capacity and congestion level. Moveover, ISPs might be able
to use the CP-side revenue to subsidize the service fees for consumers
so as to increase market share.

We envision that the Public Option should be implemented as the safety
net, or the last/back-up choice, for the consumers if the existing
commercial ISPs' strategy hurt consumer surplus. The more ISPs
competing in a market, the less the market needs a public option
and the less capacity we need to deploy for the Public Option ISP to
be effective. In the most hostile case where only one monopolistic ISP
exists in the market, a Public Option ISP could be effective as long
as it has a capacity that is larger than the percentage of consumers
that the monopoly cannot afford to lose. For example, if $10\%$ of the
market share is critical for the monopoly, implementing $10\%$ of its
capacity would be able to at least ``steal'' $10\%$ of consumers from
the monopoly if it follows a network neutral strategy. If the monopoly applies
a worse than neutral strategy for consumer surplus, it will lose even more. In that
sense, although $10\%$ of the capacity will not be operating optimally,
its existence incentivizes the remaining $90\%$
maximizing for consumer surplus, which could result in much better
consumer surplus than requiring the monopoly to follow network neutral
regulations.


Summarizing, we believe our paper sheds new light on and informs the
continuing debate on the role of regulation on the Internet and our
introduction of the Public Option ISP is an important contribution.


\appendix

\section{Additional Experiment}\label{sec:additional_experiment}
\begin{figure}[!ht]
\centering
\includegraphics[width=2.4in, angle=0]{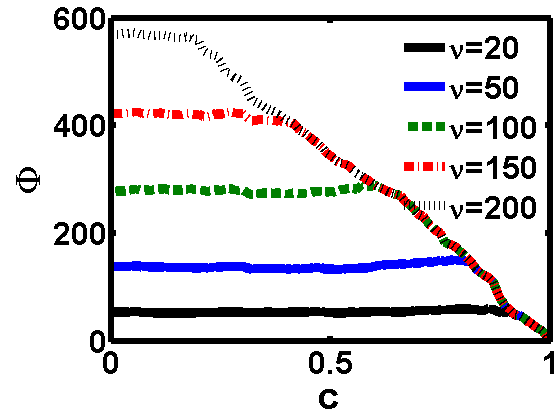}
\caption{Per capita surplus $\Phi$ under $\kappa=1$.}
\label{figure:uniform_kappa}
\end{figure}

\begin{figure}[!ht]
\centering
\includegraphics[width=7.2in, angle=0]{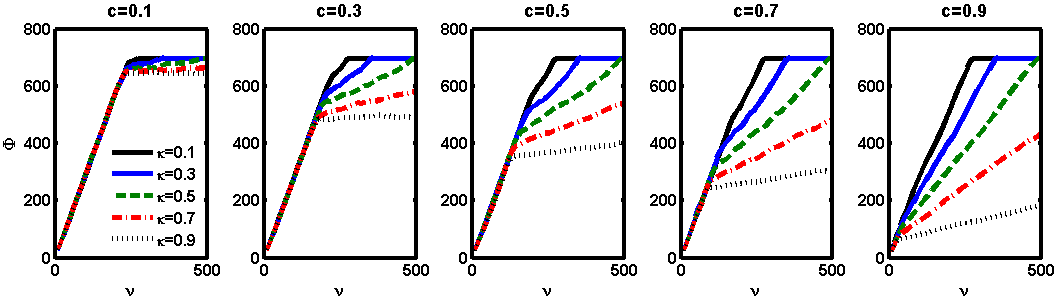}
\caption{$\Phi$ under various strategies $s_{\mathit I}=(\kappa,c)$ versus per capita capacity $\nu$.}
\label{figure:uniform_monopoly}
\end{figure}

We also perform another set of experiments of the same set of $1000$ CPs used in our experiments in Section \ref{sec:monopoly} and \ref{sec:oligopoly}. The only difference is that the parameter $\phi_i$ has a distribution $\phi_i = U[0,U[0,10]]$, which is independent of $\beta_i$. Figure \ref{figure:uniform_kappa} and \ref{figure:uniform_monopoly} illustrate the consumer surplus under $\kappa=1$ and various strategies of the ISP respectively. Since the characteristics of the CPs are the same as our previous experiments, the CPs' decision and the ISP's revenues are the same as before.

\begin{figure}[!ht]
\centering
\includegraphics[width=2.3in, angle=0]{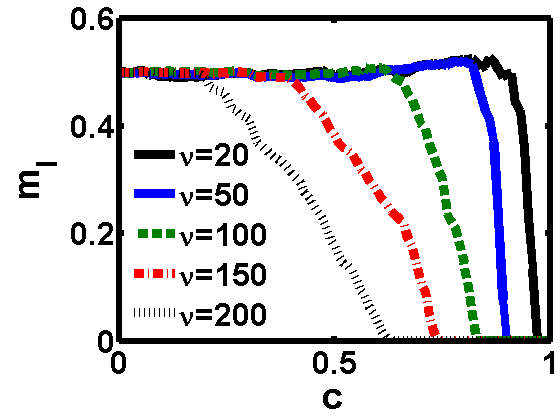}
\includegraphics[width=2.3in, angle=0]{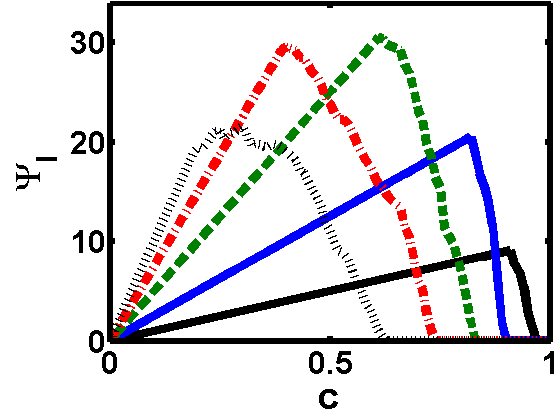}
\includegraphics[width=2.3in, angle=0]{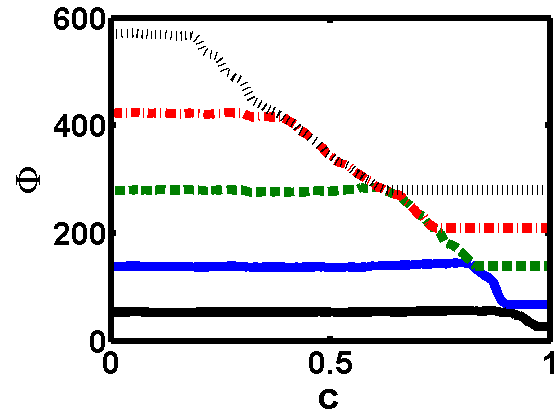}
\caption{ISP $\mathit I$'s market share $m_{\mathit I}$ and per capita surplus $\Psi_{\mathit I}$ and
per capita consumer surplus $\Phi$ under $\kappa=1$.}
\label{figure:uniform_kappa2}
\end{figure}

\begin{figure*}[!ht]
\centering
\includegraphics[width=7.2in, angle=0]{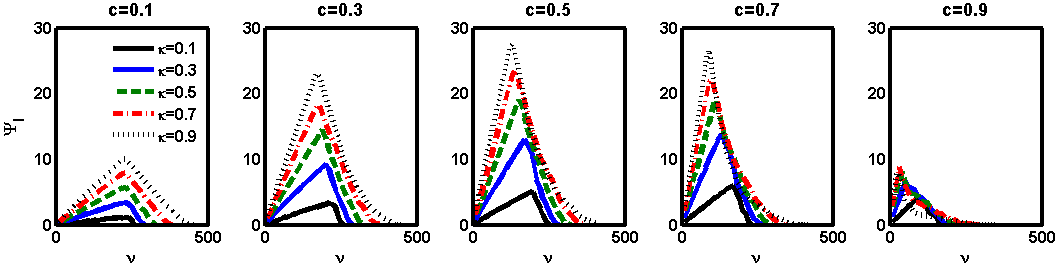}
\includegraphics[width=7.2in, angle=0]{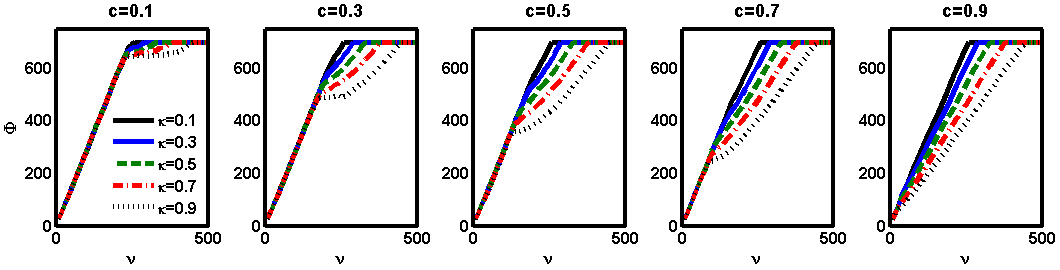}
\includegraphics[width=7.2in, angle=0]{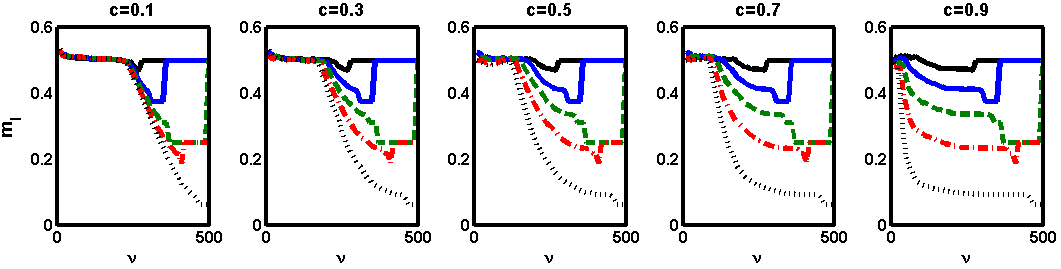} \caption{ISP $\mathit I$'s market share $m_{\mathit I}$ and per capita surplus $\Psi_{\mathit I}$ and
per capita consumer surplus $\Phi$ under $\kappa=1$.}
\label{figure:uniform_oligopoly}
\end{figure*}

Figure \ref{figure:uniform_kappa2} and \ref{figure:uniform_oligopoly} show the results that are parallel to figure \ref{figure:Kappa_Oli} and  \ref{figure:oligopoly}. We still find the same observations as in our previous experiments.

\end{document}